\documentclass[twocolumn,pre,superscriptaddress,floatfix,longbibliography]{revtex4-1}

\usepackage{graphics}
\usepackage{graphicx}
\usepackage{amsmath}
\usepackage{amssymb,xcolor}
\usepackage{gensymb}
\usepackage{ulem}

\newcommand{\textin}[1]{\mbox{\scriptsize{#1}}}

\newcommand{\blue}[1]{\textcolor{black}{#1}}

\definecolor{grisclair}{rgb}{0.6,0.6,0.6}

\newcommand{\beq}{\begin{equation}}
\newcommand{\ee}{\end{equation}}

\begin{document}

\title{A new emitter for electrospray and electrohydrodynamic jet printing}
\author{D. Fern\'andez-Mart\'{\i}nez}
\address{Depto.\ de Ingenier\'{\i}a Mec\'anica, Energ\'etica y de los Materiales and\\ 
Instituto de Computaci\'on Cient\'{\i}fica Avanzada (ICCAEx),\\
Universidad de Extremadura, E-06006 Badajoz, Spain}
\author{E. J. Vega}
\address{Depto.\ de Ingenier\'{\i}a Mec\'anica, Energ\'etica y de los Materiales and\\ 
Instituto de Computaci\'on Cient\'{\i}fica Avanzada (ICCAEx),\\
Universidad de Extremadura, E-06006 Badajoz, Spain}
\author{A. M. Ga\~n\'an-Calvo}
\address{Departamento de Ingenier\'{\i}a Aeroespacial y Mec\'anica de Fluidos,\\
Universidad de Sevilla, E-41092 Sevilla, Spain}
\author{J. M. Montanero}
\email[Corresponding author: J.M. Montanero]{(jmm@unex.es)}
\address{Depto.\ de Ingenier\'{\i}a Mec\'anica, Energ\'etica y de los Materiales and\\ 
Instituto de Computaci\'on Cient\'{\i}fica Avanzada (ICCAEx),\\
Universidad de Extremadura, E-06006 Badajoz, Spain}

\begin{abstract}
We propose using a dielectric beveled nozzle for electrospray and electrohydrodynamic jet printing. This nozzle stabilizes the liquid ejection of low-conductivity liquids, considerably reducing the minimum flow rate below which the flow becomes unstable. This translates into a significant reduction of the minimum jet diameter. Due to its dielectric character, electrochemical reactions occurring in metallic beveled nozzles (e.g. hypodermic needles) do not occur, preserving the purity of the liquid. This property makes this nozzle appropriate for Electrospray Ionization Mass Spectrometry (ESI-MS) or bioplotting. We illustrate the capabilities of this new technique by conducting (i) electrospray experiments with Newtonian liquids and (ii) electrohydrodynamic jet printing experiments with viscoelastic fluids. Jets with diameters around 1 $\mu$m are produced with low-conductivity liquids such as octanol and glycerine. Viscoelastic threads a few microns in diameter are gently deposited on a moving substrate to print out uniform lines tens of nanometers in height. Due to the strong stabilizing effect of the beveled nozzle, the minimum flow rate and jet diameter were much smaller than the respective values obtained with the cylindrical capillary in the electrospray and electrohydrodynamic jet printing experiments. The proposed technique opens new routes for electrospray and electrohydrodynamic jet printing.
\end{abstract}

\maketitle

\section{Introduction}

Under certain conditions, the interfacial Maxwell stress produced by an intense electric field accelerates the liquid emanating from a cylindrical feeding capillary. The fluid meniscus attached to the feeding capillary adopts a conical shape and steadily emits an electrified jet from its tip \citep{GLHRM18,RGL18}. The electric stress can reduce the jet diameter several orders of magnitude below the size of the feeding capillary, thus avoiding clogging effects. 

In the classical cone-jet mode of electrospray, the jet diameter $d_j$ scales as $d_j\sim d_0 (Q/Q_0)^{1/2}$, where $d_0=[\sigma\varepsilon_0^2/(\rho K^2)]^{1/3}$, $Q$ is the injected flow rate, and $Q_0=\sigma \varepsilon_o/(\rho K)$ \citep{G04a}. Here, $\sigma$ is the surface tension, $\varepsilon_0$ is the vacuum dielectric constant, $\rho$ is the liquid density, and $K$ is the electrical conductivity. As can be seen, the jet diameter $d_j$ scales as the square root of the injected flow rate $Q$. This implies that the simplest way of reducing the jet diameter is to decrease the flow rate. However, there is a minimum flow rate $Q_{\textin{min}}$ below which the cone-jet mode becomes unstable. Dimensional analysis dictates that the minimum flow rate obeys the formal relationship $Q_{\textin{min}}/Q_0=f(\varepsilon,\delta_{\mu})$ as long as the stability of the cone-jet mode essentially relies on the local flow in the cone-jet transition region. Here, $\delta_{\mu}=[\sigma^2\rho\varepsilon_0/(\mu^3 K)]^{1/3}$ is the electrohydrodynamic Reynolds number \citep{G04a}. It is natural to hypothesize that $Q_{\textin{min}}/Q_0\simeq g(\varepsilon)$ for $\varepsilon\gg \delta_{\mu}^{-1}$ and $Q_{\textin{min}}/Q_0\simeq h(\delta_{\mu})$ for $\varepsilon\lesssim \delta_{\mu}^{-1}$. Experimental results are consistent with the simple scaling laws $Q_{\textin{min}}/Q_0\sim \varepsilon$ and $Q_{\textin{min}}/Q_0\sim \delta_{\mu}^{-1}$ for the polar and viscous limits, respectively \citep{GRM13}. 

The above results imply that the minimum jet diameter is at least tens of microns for low-conductivity liquids, such as many alcohols and glycerine. This may constitute an important limitation in several applications. A natural question is whether replacing the cylindrical feeding capillary with a different injector can significantly lower the minimum flow rate stability limit and lead to much smaller emitted jets.

Different alternatives to the cylindrical feeding capillary have been proposed for several purposes. For instance, externally wetted emitters are used in the electrospray of ionic liquids to develop ion engines for space missions. Electrochemically sharpened and roughened tungsten wires with a few microns tip radii are commonly used to propel micro–satellites \citep{LM05,LM04,CF09}. These emitters have also been fabricated on a Si wafer \citep{VAM06} by a MEMS process \citep{NTT17}. Emitters microfabricated in porous metals \citep{LL11,HLLSXF22} and borosilicate glass \citep{CS15} have also been utilized for this purpose. 

The development of low-flow rate emitters for Electrospray Ionisation Mass Spectrometry (ESI-MS) \citep{FMMWW89} has also improved the efficiency of this technique in terms of the amount of sample required. Pointed carbon fiber emitters \citep{LRBK04,SDKL06} constitute a good example of this. Micrometer emitters made of borosilicate and fused silica are commonly used in nanoelectrospray for highly sensitive mass spectrometry in biological research \citep{SKD03,GMO09}. 

Electrospray has been used in many other areas. For instance, it can be applied to sputter deposition in surface coating. Nano-droplets were produced to this end by a specially designed wire needle valve \citep{S99}.

Electrohydrodynamic (EHD) jet printing is an additive manufacturing technique that can produce high-resolution micro patterns \citep{RKMJC12,YKSKACC18}, organic field-effect transistors \citep{LJJLK18}, thin-film transistors \citep{CKC21}, and photodetectors sensors \citep{HBCCK22,BWHWTSJHY23,HSR24}, among other microelectronic products. EHD jet printing is also used to extrude continuous submillimeter filaments of ``bioinks" consisting of aqueous media, thermoreversible polymers, or polymer/hydrogel precursors combined with living cells \citep{ROBYS06}.

Ultra-high-resolution EHD jet printing was achieved in the pioneering work of \citet{Parketal} with a gold-coated glass microcapillary nozzle with an internal diameter of 2 $\mu$m. Nozzles with diameters down to 100 nm can be formed by pulling glass pipettes. However, these nozzles suffer from clogging. \citet{L05} and \citet{KKPH10} proposed using a polymeric tip coated with non-conductive resin and coaxially inserted in a cylindrical stainless steel nozzle. This procedure was demonstrated to stabilize the jet emission of Newtonian liquids, reducing the minimum voltage to produce that emission. \citet{WQPSZLS12} conducted experiments with a similar device and viscoelastic liquids.

The emitters described above entail a drastic reduction in the size of the passages crossed by the liquid to reach the Taylor cone. This inevitably produces clogging effects, especially when particles and macromolecules are dissolved in the liquid. To avoid this, \blue{conducting} beveled nozzles (also frequently referred to as tilted nozzles) have been proposed as valuable alternatives to the cylindrical capillary in electrospray, EHD jet printing, and other tip streaming configurations. For instance, beveled nozzles have been used in flow focusing to produce microemulsions \citep{ARMGV13}, microdroplets \citep{RAMMG16}, and microjets \citep{NZAKEMNSWAK20} for serial femtosecond crystallography. 

\blue{Conducting} beveled nozzles have also been utilized in electrospray for several purposes. Nanostructured thin ceramic \citep{CEKMS99,KPCWT08} and calcium phosphate \citep{LWSJ03} films have been deposited with a beveled stainless steel nozzle \citep{JSK18}. Fine and highly conductive patterns of silver ink have been fabricated by near-field EHD jet printing with this device \citep{YKYLKASRSY09,KAO15}. A resolution of less than 2 $\mu$m was achieved with hypodermic needles of diameters larger than 250 $\mu$m \citep{RA21}. \citet{Vuetal22} showed the advantages of the chamfered nozzle in fabricating highly uniform polymeric fibers for piezoelectric sensor development. 

Beveled nozzles are known to stabilize the jet emission, reducing the minimum flow rate below which the cone-jet mode becomes unstable and, therefore, the size of the ejected filament and droplets. However, the electric field strength reaches very high values in the tip of the metallic nozzle, which may produce undesirable electrochemical reactions not present with the flat (cylindrical) feeding capillary. Those reactions could inject positive ions into the solution, significantly altering its chemical composition \citep{RMG13}. A similar effect can be observed in micrometer metallic emitters of electrospray. 

\begin{figure}[hbt]
\begin{center}
\resizebox{0.425\textwidth}{!}{\includegraphics{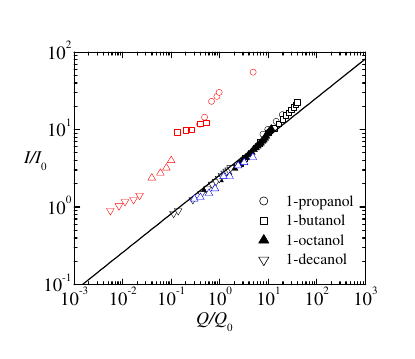}}
\end{center}
\caption{Electric current $I/I_0$ as a function of the flow rate $Q/Q_0$ \citep{RMG13}. The red and black symbols correspond to the beveled needle and cylindrical capillary configurations, respectively. The blue symbols correspond to the dielectric beveled nozzle proposed in this work. The solid line is the scaling law $I/I_0=2.6 (Q/Q_0)^{1/2}$ \citep{G04a}.} 
\label{fig0}
\end{figure}

\begin{figure}[hbt]
\begin{center}
\resizebox{0.465\textwidth}{!}{\includegraphics{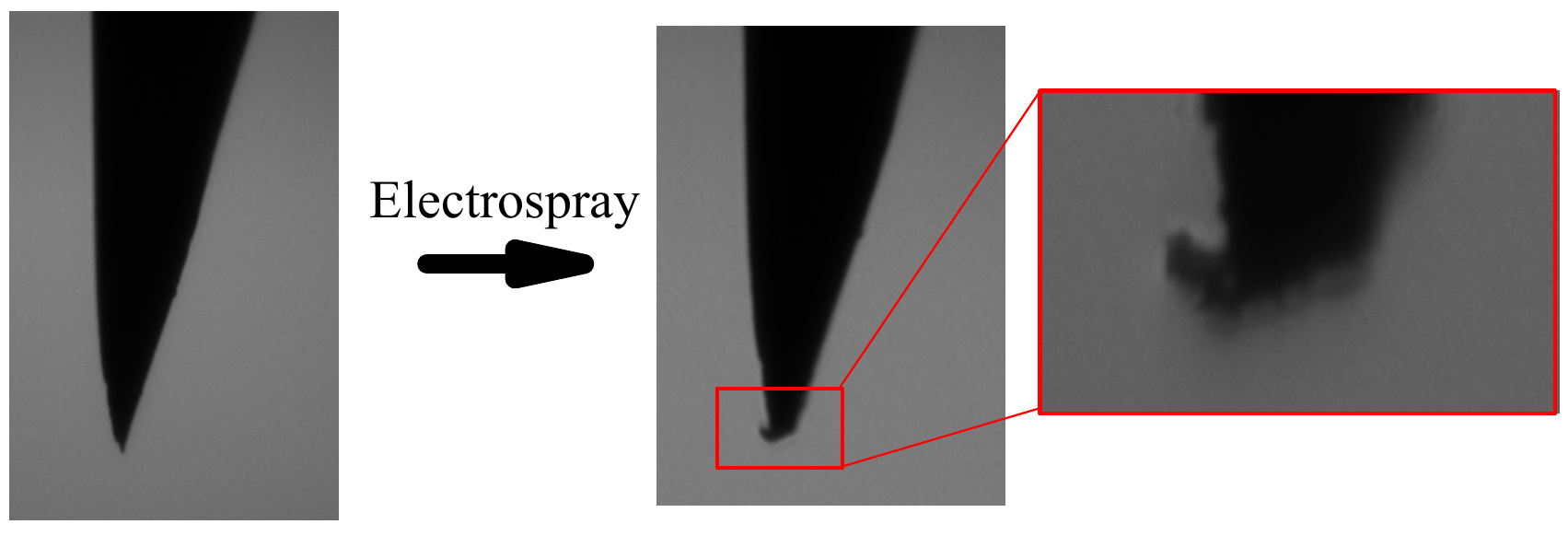}}
\end{center}
\caption{Images of the hypodermic needle before and after the electrospray experiment \citep{RMG13}.} 
\label{fig00}
\end{figure}

Figure \ref{fig0} shows the electric current $I$ in terms of its characteristic value $I_0=\varepsilon_0^{1/2}\sigma \rho^{-1/2}$ as a function of the flow rate $Q$ relative to its characteristic value $Q_0$ defined above \citep{G04a}. The electric current with the hypodermic needle was one order of magnitude larger than that with the cylindrical capillary. \citet{RMG13} attributed this discrepancy to an increase in the liquid electrical conductivity of up to two orders of magnitude due to local electrochemical reactions at the needle tip. In fact, the tip images taken after the experiments show the loss of metallic material owing to those reactions (see Fig.\ \ref{fig00}). For this reason, beveled metallic nozzles are inadequate for many applications, such as ESI-MS and bioplotting.

We conclude that there is no satisfactory alternative to the cylindrical capillary to stabilize the ejection of low-conductivity liquids without the risk of clogging and undesired electrochemical reactions in the electrode. To fulfill this triple condition, we propose using a {\em dielectric} beveled nozzle that can be applied to both electrospray and electrohydrodynamic jet printing. 

This work analyzes the electrospray of Newtonian fluids and EHD jet printing of viscoelastic liquids with a beveled dielectric nozzle. The nozzle is made of IP-S resin, a biocompatible material that does not suffer from degradation due to the electric field or swelling caused by the liquids used in our experiments. We show the large stabilizing effect of the beveled nozzle, which allows one to produce liquid jets at flow rates much lower than those in the equivalent cylindrical capillary configuration. We study the dependence of the fluid jet diameter on the flow rate in both the electrospray and EHD jet printing applications. We analyze the current intensity transported by the electrosprayed liquid and the morphology of the printed lines. These results shows that this novel technique has broad applicability across diverse fields such us ESI-MS, ion thursters, and additive manufacturing.

\section{Methods}

\subsection{Nozzle fabrication}

\begin{figure*}[hbt]
\begin{center}
\resizebox{0.6\textwidth}{!}{\includegraphics{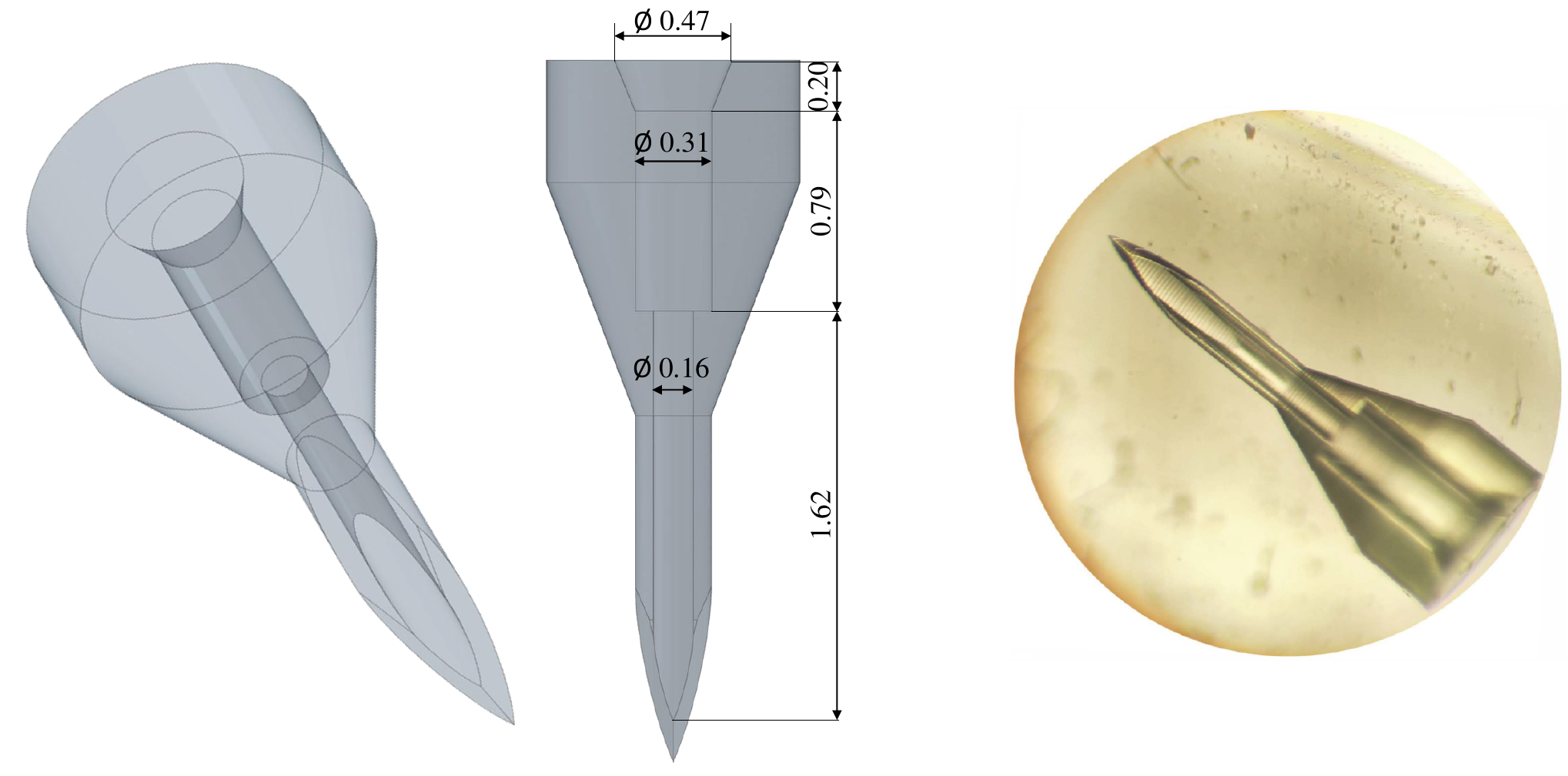}}
\end{center}
\caption{Design and image of the nozzle used in the experiments (lengths measured in mm).} 
\label{nozzle}
\end{figure*}

The beveled nozzle (Fig.\ \ref{nozzle}) was printed using Nanoscribe Photonic Professional GT2 with the Dip-in Laser Lithography (DiLL) configuration, dipping the 25$\times$ objective into the biocompatible IP-S resin droplet on an ITO-coated glass substrate. We chose the shell and scaffolds writing strategy. We produced a 20 $\mu$m thickness shell to delimit the structure and an internal scaffold to stabilize it. The typical slicing and hatching distances were $1$ $\mu$m and 0.5 $\mu$m. The part was developed in $\sim$25 ml of propylene glycol monomethyl ether acetate (PG-MEA) for 24 h and then cleaned in isopropanol for 2 h. Then, unexposed resin inside the shell was cured for 60 min inside the UV Curing Chamber (XYZprinting).

The design of our beveled nozzle is a reproduction in resin of the stainless steel hypodermic needle from Fig.\ \ref{fig00} (Becton Dickinson Microlance 3 30G 1/2) with an inner (outer) diameter of about 160 (300) $\mu$m. The tube ends in a beveled tip with an outer hydraulic radius of a few microns (the tip curvature radius is around 2.5 $\mu$m). It must be pointed out that the tube size is similar to that of the cylindrical feeding capillaries typically used in electrospray and EHD jet printing experiments, which eliminates clogging effects. The nozzle was connected to the end of a standard stainless steel capillary (30G) under pressure.

\subsection{Experimental setup}

In the electrospray experiments, we injected the liquid at a constant flow rate $Q$ with a syringe pump (KD Scientific, Legato 210 Series) connected to a stepping motor. The liquid exits through the beveled nozzle (A). The metallic plate had an orifice (B) of 200 $\mu$m in diameter in front of the needle. The plate covered the upper face of a metallic cubic cell (C). We used a high-precision orientation system (D) and a translation stage (E) to ensure the correct alignment of these elements and to set the capillary-to-orifice distance. 

An electric potential $V$ was applied to the metallic part of our feeding capillary (30G) through a DC high-voltage power supply (Bertan 205B-10R) (F). The cubic cell was used as the ground electrode. The feeding capillary polarity was chosen positive. A prescribed negative gauge pressure (about 20 mbar) was applied in the cubic cell using a suction pump (Hanning Elektro-Werke SV 1003 D000) (G) to produce an air stream coflowing with the jet. The liquid jet and the coaxial air stream crossed the plate orifice, preventing liquid accumulation on the metallic plate. The electric current $I$ transported by the liquid jet was measured using a picoamperimeter (9103 HSPR, rdb instruments) (H) connected to the cell. Special care was taken for electrical shielding and grounding because the full scale of typical electric current measurements was in the nanoampere range. It is worth noting that the force exerted by the air stream is much smaller than the electric one. Therefore, the air stream does not affect the jet dynamics. 

\begin{figure}[!]
\begin{center}
\resizebox{1.0\columnwidth}{!}{\includegraphics{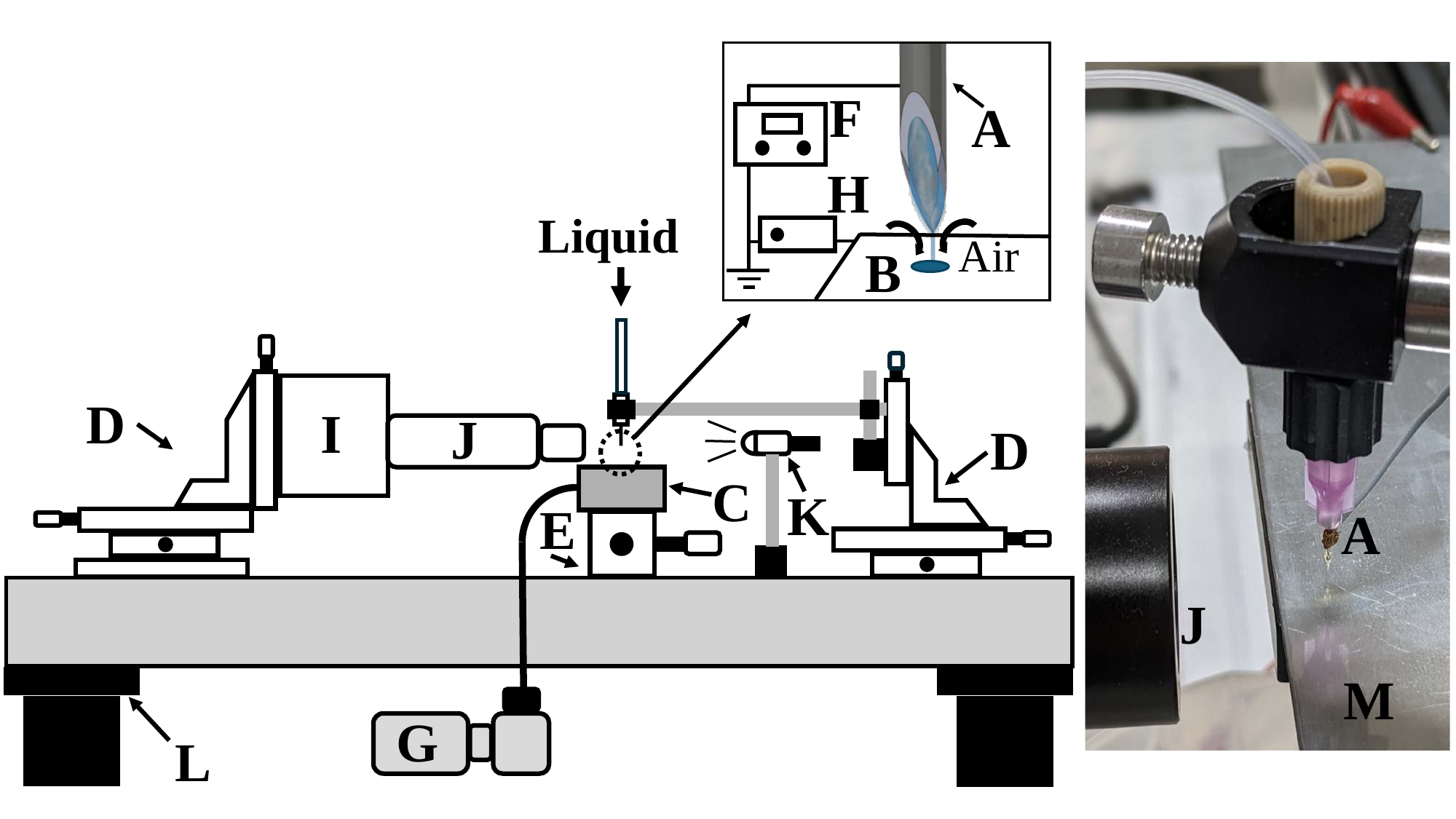}}
\end{center}
\caption{Experimental setup: (A) needle, (B) plate with orifice, (C) suction cell, (D) orientation systems, (E) translation stage, (F) high voltage power supply,  (G) suction pump, (H) picoamperimeter, (I) camera, (J) optical lenses, (K) optical fiber connected to a light source, and (L) anti-vibration isolation system.}
\label{Exp_Sketch}
\end{figure}

Digital images of the liquid jets were acquired at 2000 frames per second with an exposure time in the interval $3.9-10$ $\mu$s using a high-speed camera (Photron FastCam Mini UX100) (I) equipped with optical lenses (Optem Zoom 70XL) (J). The magnification was adjusted within the range $0.20-1.33$ $\mu$m/pixel in each experiment. The fluid configuration was illuminated from the back by a halogen light source (Leica KL 2500 LCD) (K). All these elements were mounted on an optical table with a pneumatic anti-vibration isolation system (L) to dampen the vibrations coming from the building.

In EHD jet printing experiments, the viscoelastic jet was deposited on a cover glass with a thickness in the interval $0.13-0.17$ mm (RS France). The cover glass was hydrophilic and its roughness was in the order of a few nanometers. The cubic cell utilized in the cone-jet mode experiments was removed. The cover glass was positioned above a metallic plate located on a horizontal platform (M). The platform was displaced horizontally at a constant velocity $v_s=0.46$ m/s using a belt-driven linear actuator coupled to a DC motor.

\subsection{Liquids}

The electrospray experiments were conducted with 1-octanol and glycerine. Octanol is a leaky-dielectric (low-conductivity) liquid commonly used in electrospray. The instability of the cone-jet mode at the minimum flow rate of octanol is caused by the polarization force opposing the flow in the cone-jet transition region \citep{GRM13}. The glycerine viscosity is two orders of magnitude larger than that of octanol (Table \ref{tab1}). In this case, the instability is caused by the viscous force. Using octanol and glycerine allows us to assess the performance of our emitter in the electrospray polarization and viscous regimes.

\begin{table*}
\begin{tabular}{|c|c|c|c|c|c|c|c|}
\hline
&$\rho$ (kg/m$^3)$ & $\mu$ (kg/m$\cdot$s) & $\sigma$ (mN/m) & $\lambda$ (ms) & $K$ ($\mu$S/m) & $\varepsilon$ & $\delta_{\mu}$ \\
\hline
octanol & 827 & 0.0072 & 23.5 & 0 & 0.9 & 10 & 2.2918 \\
\hline
glycerine & 1261 & 1.2 & 62.2 & 0 & 1 & 42.5 & 0.0292 \\
\hline
octanol-PVP 1\% & 828 & 0.0315 & 22 & 1.53 & 4 & 10 & 0.3187 \\
\hline
octanol-PVP 5\% & 839 & 0.5649 & 20.9 & 8.40 & 4 & 10 & 0.0178 \\
\hline
\end{tabular}
\caption{Liquid density $\rho$, zero-shear rate viscosity $\mu$, surface tension $\sigma$, extensional relaxation time $\lambda$, electrical conductivity $K$, relative permittivity $\varepsilon$, and electrohydrodynamic Reynolds number $\delta_{\mu}=[\sigma^2\rho\varepsilon_0/(\mu^3 K)]^{1/3}$. The surface tension, the extensional relaxation time, and electrical conductivity were measured with the TIFA method \citep{CBMN04}, the slow retraction method \citep{SVSMA17}, and the procedure described by \citet{PRHGM18}, respectively.}
\label{tab1}
\end{table*}

\begin{figure}[!]
\begin{center}
\resizebox{0.7\columnwidth}{!}{\includegraphics{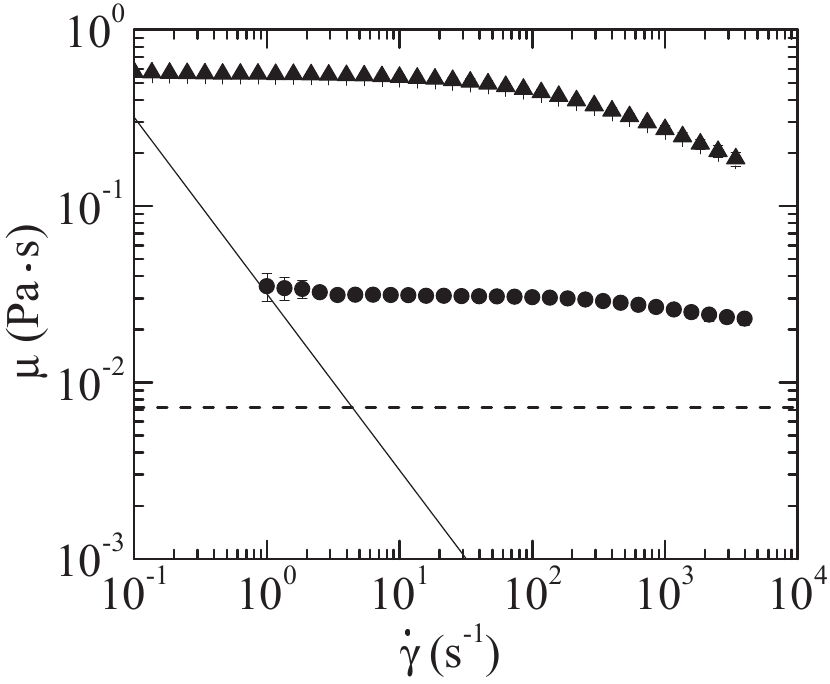}}
\end{center}
\caption{Shear viscosity $\mu$ as a function of the shear rate $\dot{\gamma}$. The dashed horizontal line represents the shear viscosity of the solvent. The solid line indicates the minimum measurable shear viscosity based on 80$\times$ the minimum resolvable torque of the shear rheometer (Kinexus Pro+ Rheometer, plate-plate geometry PU40 SR6002 SS, and a gap of 200 $\mu$m).}
\label{shear}
\end{figure}

We dissolved Polyvinylpyrrolidone (PVP) in octanol at the mass concentrations $1\%$ and 5\% to conduct EHD jet printing experiments. Hereafter, we will refer to these two polymer solutions as PVP 1\% and PVP 5\%. PVP is a biocompatible polymer frequently used in bioplotting applications. The conductivities of these viscoelastic solutions are larger than those of octanol owing to the ions released during the polymer dissolution. PVP 1\% and PVP 5\% exhibit a slight shear-thinning in the range of shear rates analyzed in our shear-rheology experiments (Fig.\ \ref{shear}). One expects significant extensional viscoelastic effects on the flow occurring in the cone-jet transition region owing to the magnitude of extensional relaxation time (Table \ref{tab1}).

\subsection{Experimental procedure}

\begin{figure}[!]
\begin{center}
\resizebox{0.95\columnwidth}{!}{\includegraphics{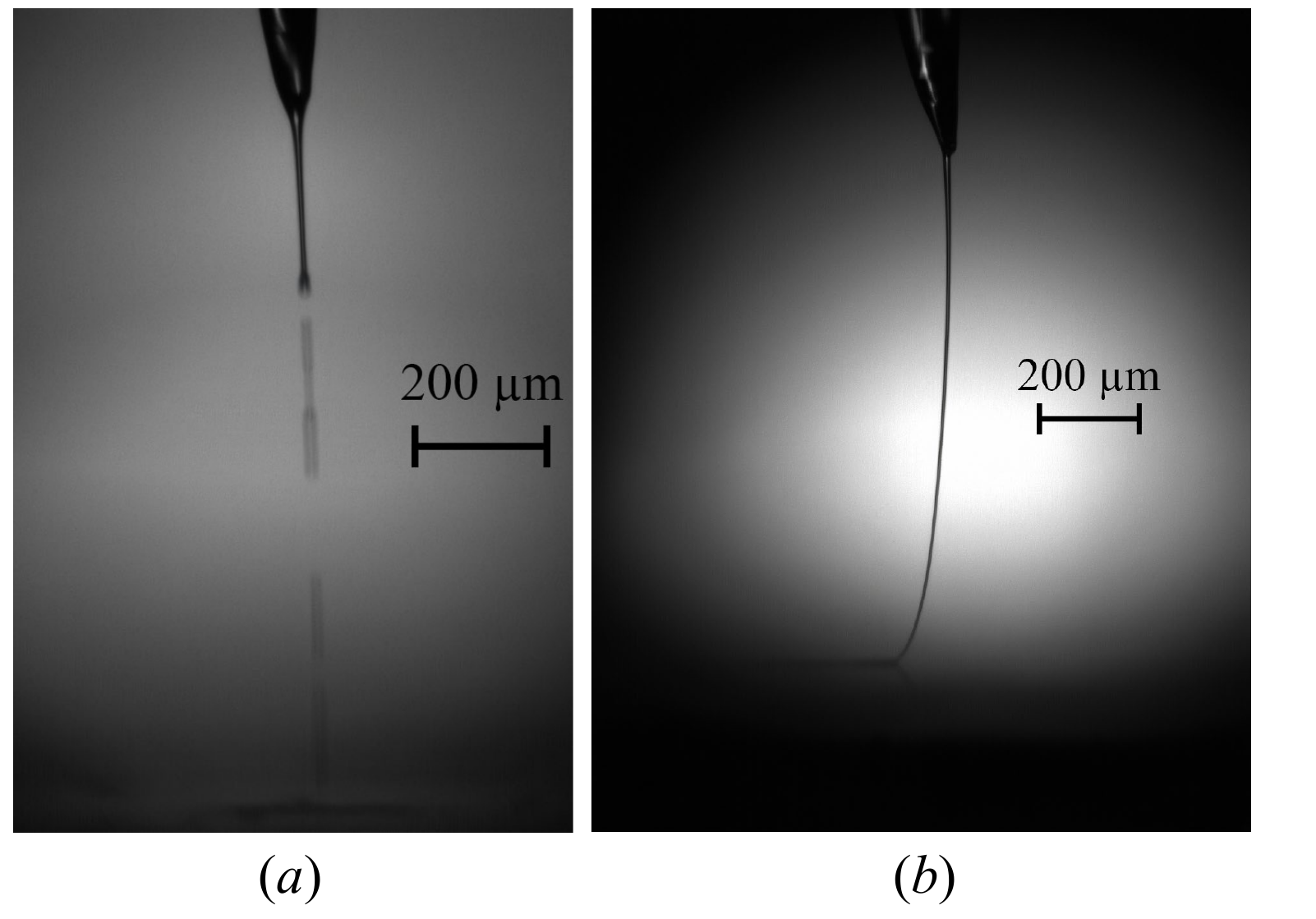}}
\end{center}
\caption{(a) Image of the cone-jet mode of electrospray for octanol and $Q=0.8$ ml/h. (b) Image of the printing for EHD jet printing for 1\% PVP and $Q=0.2$ ml/h.}
\label{ilus}
\end{figure}

\begin{figure}[!]
\begin{center}
\resizebox{0.65\columnwidth}{!}{\includegraphics{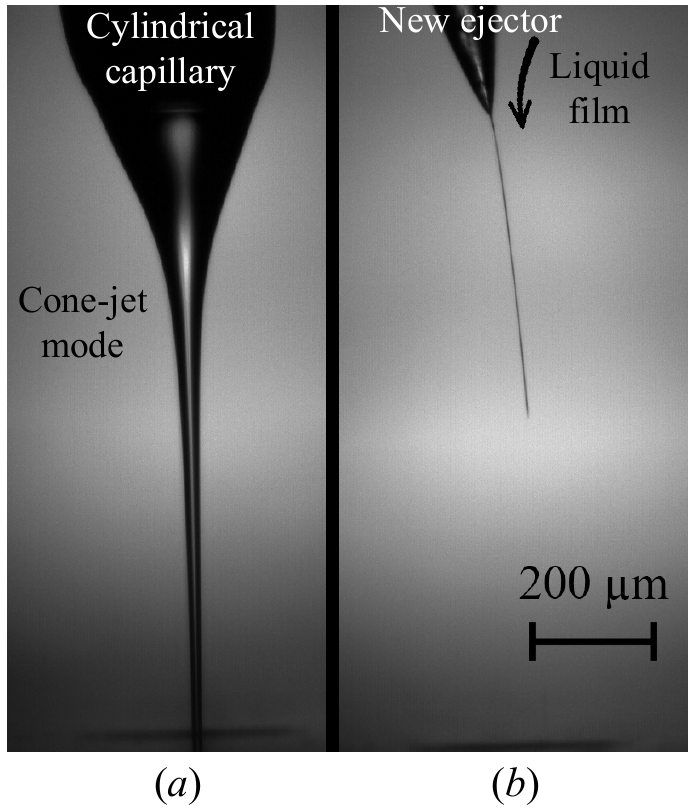}}
\end{center}
\caption{\blue{Image of the cone-jet mode of electrospray (a) and the liquid film on the emitter tip (b) for glycerine, $H=1$ mm, $V=2$ kV, and the corresponding minimum flow rates $Q=2.5$ ml/h (a) and 0.004 ml/h (b).}}
\label{ilus2}
\end{figure}

For the sake of illustration, Fig.\ \ref{ilus} shows images of the steady jetting mode of electrospray and EHD jet printing with our beveled nozzle. In this configuration, the conical meniscus is replaced by a much more stable liquid film. The film flows over the nozzle's outer surface towards the tip, driven by the electric forces. The jet is emitted from a tiny Taylor cone formed in the needle tip (Fig.\ \ref{ilus2}). In electrospray experiments, the electrified jet of the Newtonian liquid breaks up into charged droplets. In EHD jet printing, polymers stretch in the emission point, sharply increasing the liquid extensional viscosity. The jet flies during a time shorter than or of the order of the extensional relaxation time, and the extensional viscosity does not relax to the value corresponding to the polymer coiling state. This allows us to deposit the ejected filament on the substrate before instability develops.

The images shown in Fig.\ \ref{ilus} correspond to flow rates much larger than the minimum flow rate stability limit $Q_{\textin{min}}$. In our experiments, we fixed the emitter-to-counter-electrode distance $H$ and the applied voltage $V$. Then, we progressively decreased the flow rate to $Q_{\textin{min}}$, at which the steady ejection ceased. We verified that the instantaneous electric current measured in the electrospray experiments was constant for $Q>Q_{\textin{min}}$.  

The images were analyzed with an in-house Python script to determine the jet diameter $d_j$. The morphology of the printed lines in the EHD jet printing experiments was analyzed with a scanning electron microscope (SEM) (TENEO, FEI) and a non-contact 3D optical profiler (Sensofar S-NEOX). Using interferometry, the optical profiler allowed us to measure the surface height of very smooth and continuous surfaces with sub-nanometer resolution. SensoMap and SensoScan software were used to process the results from the profiler.

\section{Results}

\subsection{Electrospray}

In this section, we analyze whether using the beveled nozzle significantly reduces the minimum jet diameter obtained in the classical electrospray configuration, i.e., that in which the liquid is injected with the standard cylindrical capillary with similar dimensions. For this purpose, we consider octanol and glycerine, for which $\varepsilon\gg \delta_{\mu}^{-1}$ and $\varepsilon\simeq \delta_{\mu}^{-1}$, respectively. As explained in the Introduction, this implies that the minimum flow rate is imposed by the polarization and viscous forces, respectively.

First, we compare our results for octanol with those previously obtained with the standard cylindrical capillary \citep{PRHGM18}. The applied voltage $V=2$ kV and the nozzle-to-counter-electrode distance $H=1$ mm were similar in the two experimental runs. As mentioned above, experiments were conducted varying the flow rate $Q$ to its corresponding minimum value $Q_{\textin{min}}$.

\begin{figure}[!]
\begin{center}
\resizebox{0.400\textwidth}{!}{\includegraphics{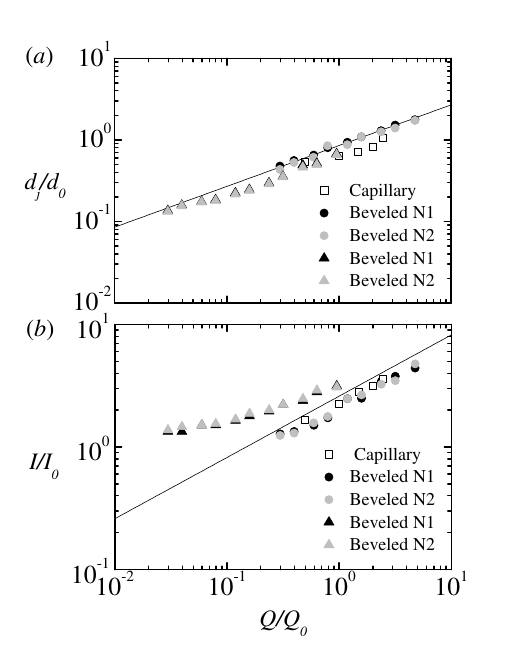}}
\end{center}
\caption{Jet diameter $d_j/d_0$ (a) and electric current $I/I_0$ (b) as a function of the flow rate $Q/Q_0$ of octanol. The circles correspond to the results obtained for $V=2$ kV and $H=1$ mm, while the triangles correspond to $V=3$ kV and $H=150$ $\mu$m. The black and gray symbols correspond to two \blue{identical} beveled nozzles, N1 and N2. The open squares correspond to the standard cylindrical capillary \citep{PRHGM18}. The lines are the scaling laws $d_j/d_0=0.85 (Q/Q_0)^{1/2}$ and $I=2.6 (Q/Q_0)^{1/2}$.}
\label{compara}
\end{figure}

As observed in Fig.\ \ref{compara}, we obtained results for $d_j$ and $I$ similar to those of \citet{PRHGM18}. Our results also follow the scaling laws for electrospray \citep{G04a}. The beveled tip does not significantly alter the outcome of the electrospray cone-jet mode. This constitutes a fundamental difference with respect to the hypodermic metallic needle used in previous works, which degrades and significantly alters the liquid chemical composition. 

Using a beveled tip allows the liquid to form a tiny Taylor cone a few micrometers in size right at the nozzle tip. The smallness of this cone and the dielectric character of the nozzle enables a drastic reduction of the nozzle-to-counter-electrode distance $H$ (this cannot be done with metallic emitters under normal ambient conditions due to the appearance of spark caused by air ionization). This miniaturization of the electrospray cone-jet mode reduces the minimum flow rate and, therefore, the minimum jet diameter (Fig.\ \ref{compara}). It must be pointed out that this reduction is achieved using a fluid passage hundreds of microns in diameter (Fig.\ \ref{nozzle}), which avoids clogging effects. In this sense, our emitter fundamentally differs from those used in nanoelectrospray \citep{SKD03,GMO09}. 

\begin{figure}[!]
\begin{center}
\resizebox{0.65\columnwidth}{!}{\includegraphics{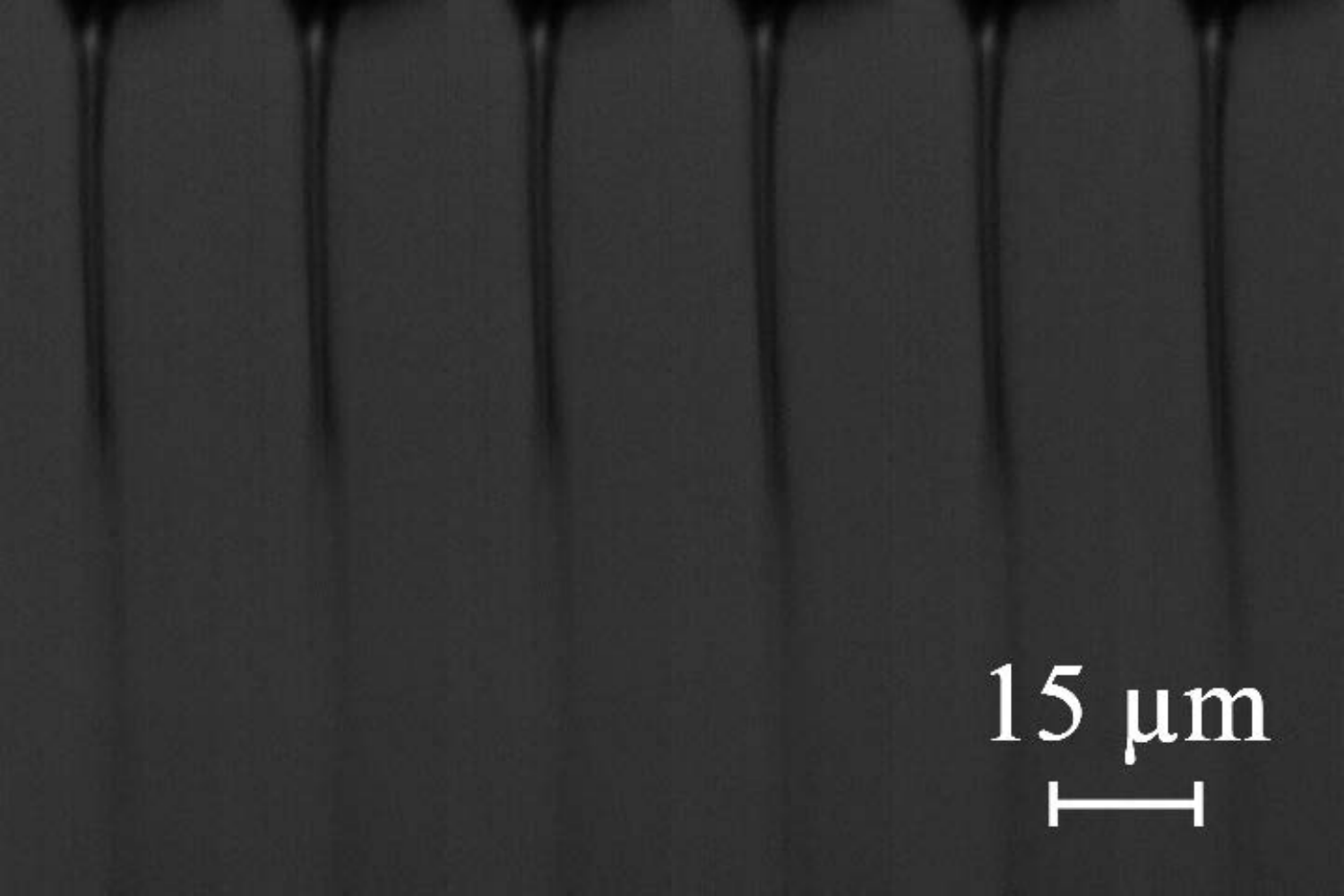}}
\end{center}
\caption{Images of the octanol jet steadily emitted with the beveled nozzle for $V=3$ kV, $H=150$ $\mu$m, and $Q=0.03$ ml/h.}
\label{nano}
\end{figure}

\begin{figure}[!]
\begin{center}
\resizebox{0.7\columnwidth}{!}{\includegraphics{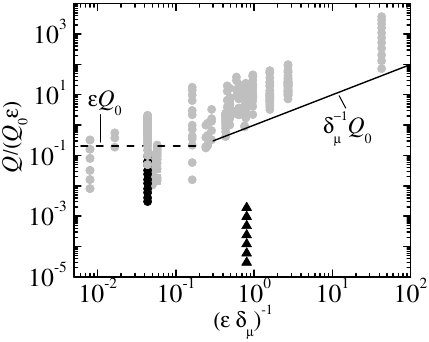}}
\end{center}
\caption{Flow rates of stable realizations for octanol with $V=3$ kV and $H=150$ $\mu$m (circles) and glycerine with $V=2$ kV and $H=1$ mm (triangles). The grey symbols correspond to previous experiments with the standard cylindrical capillary conducted by different authors \citep{PRHGM18}. The lines are the predictions obtained from the scaling analysis \citep{GRM13}.}
\label{Qmin}
\end{figure}

To illustrate the above comments, Fig.\ \ref{nano} shows an image of the jet steadily emitted with our technique by increasing the voltage to $V=3$ kV and reducing the nozzle-to-counter-electrode distance down to $H=150$ $\mu$m, which allowed us to decrease the flow rate to $Q=0.03$ ml/h in the stable regime. This value is less than $0.003\, \varepsilon\, Q_0$ ($\varepsilon\, Q_0\simeq 10$ ml/h is the scale of the minimum flow rate in the classical configuration of electrospray \citep{RMG13,FL94,H17}) and lies in the interval $Q<1000$ nl/min of the nanoelectrospray regime. Figure \ref{Qmin} shows the drastic stabilization achieved with our emitter. The jet diameter was $d_j\simeq 2$ $\mu$m. The minimum jet diameter obtained with the equivalent cylindrical-capillary configuration was $d_j=7.5$ $\mu$m with the flow rate $Q=0.5$ ml/h \citep{PRHGM18}. We did not observe any spark due to air ionization thanks to the dielectric nature of the nozzle. 

The results in Fig.\ \ref{compara} for $V=3$ kV and $H=150$ $\mu$m show that the jet diameter approximately obeys the scaling law $d_j/d_0=0.85 (Q/Q_0)^{1/2}$, which indicates that the jet velocity $v_j=4Q/(\pi d_j^2)$ is approximately the same as that measured for $H=1$ mm and that obtained with the cylindrical capillary. This means that this miniaturization of the electrospray cone-jet mode does not entail a considerable increase in viscous dissipation. Conversely, the spray current dependency on the flow rate considerably deviates from the scaling law $I\sim Q^{1/2}$. The exponent evolves from $1/2$ to $1/4$ as the flow rate decreases. This behavior has been observed in nanolectrospray (see, e.g., Ref.\ \citep{SKD03}). 

\begin{figure}[!]
\begin{center}
\resizebox{0.400\textwidth}{!}{\includegraphics{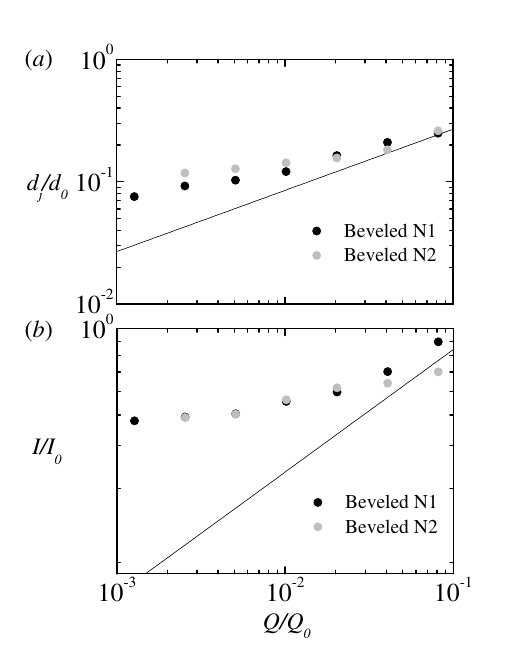}}
\end{center}
\caption{Jet diameter $d_j/d_0$ (a) and electric current $I/I_0$ (b) as a function of the flow rate $Q/Q_0$ of glicerol. The results were obtained for $V=2$ kV and $H=1$ mm. The black and gray symbols correspond to two \blue{identical} beveled nozzles, N1 and N2. The dashed lines are the scaling laws $d_j/d_0=0.85(Q/Q_0)^{1/2}$ and $I=2.6 (Q/Q_0)^{1/2}$ for the cylindrical-capillary configuration.}
\label{law}
\end{figure}

We have shown that the beveled dielectric nozzle can be used in electrospray to reduce the size of jets (droplets) of low-conductivity liquids such as octanol, for which the polar force \blue{(i.e., that caused by the permittivity jump across the interface)} sets the minimum flow rate stability limit. Now, we analyze the effect of this emitter when this limit is imposed by viscosity. To this end, we conducted experiments with glycerine using the cylindrical feeding capillary and the beveled nozzle for $V=2$ kV and $H=1$ mm in both cases.

\begin{figure}[!]
\begin{center}
\resizebox{0.95\columnwidth}{!}{\includegraphics{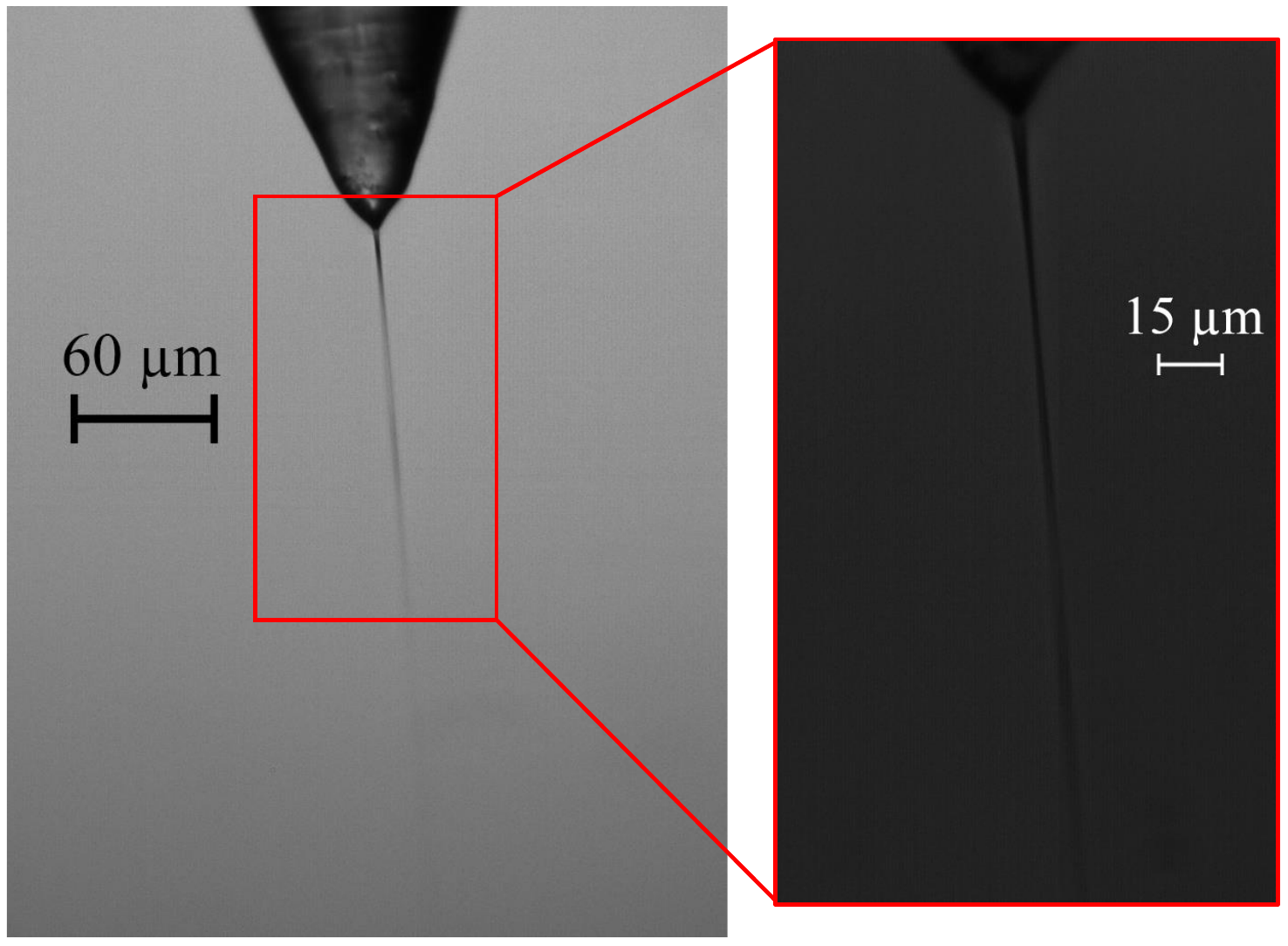}}
\end{center}
\caption{Images of the glycerine jet steadily emitted with the beveled nozzle for $V=2$ kV, $H=1$ mm, and $Q=0.002$ ml/h.}
\label{glycerine}
\end{figure}

For the cylindrical capillary, the minimum flow rate was $Q_{\textin{min}}=2.5$ ml/h, and the corresponding jet diameter was $d_j=14.9$ $\mu$m. Using the beveled nozzle led to a drastic reduction of both quantities: $Q_{\textin{min}}=0.003\pm 0.001$ ml/h and $d_j=1.5\pm 0.3$ $\mu$m (Fig.\ \ref{glycerine}). We are unaware of any electrospray realization in which such a thin glycerine jet has been steadily emitted with electrospray. \blue{Viscous friction with the beveled emitter surface stabilizes the flow, which explains why the reduction of the minimum flow rate is much larger for small electrohydrodynamic Reynolds numbers (Fig.\ \ref{Qmin})}.

The value of $Q_{\textin{min}}$ is around $5\times 10^{-5} \delta_{\mu}^{-1}\, Q_0$ ($\delta_{\mu}^{-1}\, Q_0\simeq 54$ ml/h is the scale of the minimum flow rate in the viscous regime of the classical configuration of electrospray \citep{RMG13}) (Fig.\ \ref{Qmin}). Due to viscous dissipation in the outer flow around the emitter tip, $v_j$ decreases, and $d_j$ deviates from the scaling law $d_j\sim Q^{1/2}$ as the flow rate decreases (Fig.\ \ref{compara}). The exponent of the current intensity evolves from $1/2$ to $1/4$ as the flow rate decreases. The electric current reaches an almost constant value for flow rates close to the minimum value.

The minimum flow rate instability of the classical electrospray cone-jet mode is associated with a stagnation point at the tip of the Taylor cone. As mentioned above, a liquid film slowly sliding over the ejector's outer surface replaces the Taylor cone in the present technique. This eliminates the stagnation point instability and considerably reduces the minimum flow rate.

We explored the reproducibility of our results by conducting the experiments with two \blue{identical nozzles (N1 and N2)} fabricated following the same procedure. As can be observed in Figs.\ \ref{compara} and \ref{law}, the results for octanol are practically the same. The nozzle has a slight influence on the results for glycerine. This may be due to (i) the critical role of wetting for the extremely small values of the liquid flow rate in these experiments and (ii) the fact that the jet emission region is commensurate with the printing technique resolution. 

\subsection{EHD jet printing}

The dielectric beveled nozzle can produce viscoelastic jets a few microns in diameter even when using solvents with very low conductivities, such as octanol. These jets can be deposited on a moving substrate to print polymeric lines. This section shows the capabilities of this EHD jet printing technique, considering octanol as a solvent and PVP as an example of a biocompatible polymer. We considered two polymer mass concentrations, 1\% and 5\%, corresponding to weakly and moderately viscoelastic liquids. The liquid threads are deposited on a hydrophilic dielectric surface.

\begin{figure}[hbt]
\begin{center}
\resizebox{0.4\textwidth}{!}{\includegraphics{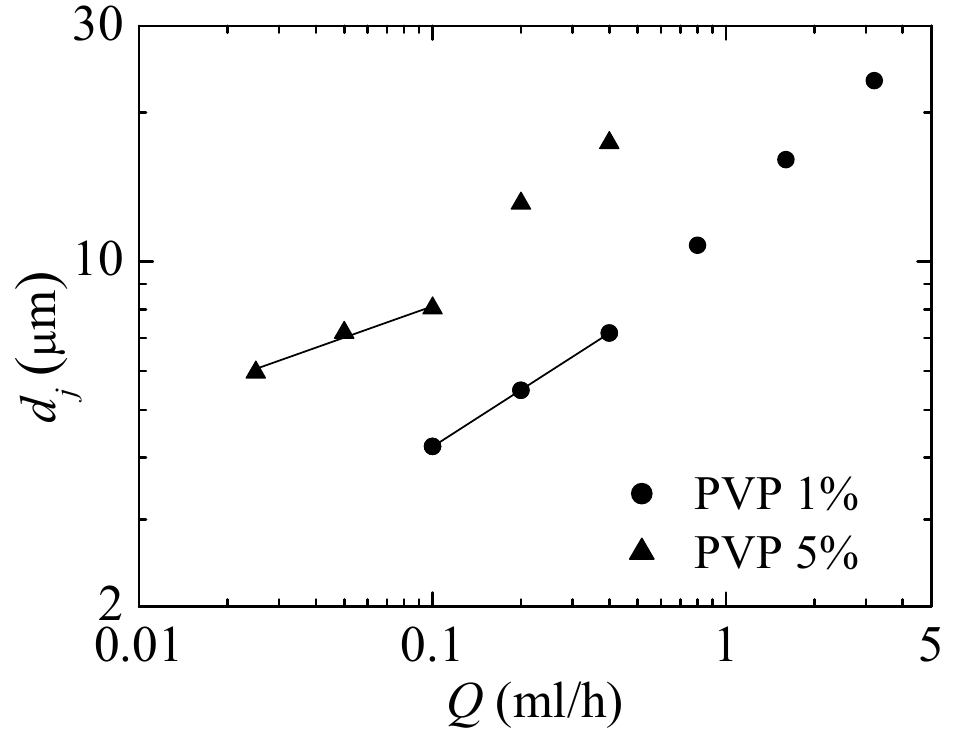}}
\end{center}
\caption{Jet diameter $d_j$ as a function of the inlet flow rate $Q$ for $V=2$ kV and $H=1$ mm. The solid lines are the fits $d_j=10.2\, Q^{0.39}$ and $d_j=13.2\, Q^{0.22}$ for PVP 1\% and PVP 5\%, respectively.} 
\label{dj-Q}
\end{figure}

Figure \ref{dj-Q} shows the jet diameter $d_j$ measured right before deposition on the substrate as a function of the flow rate $Q$ for the two polymer concentrations considered in our analysis. The flow rate was decreased to its minimum value $Q_{\textin{min}}$ for each concentration. Increasing the amount of polymer leads to an increase in the shear and extensional viscosities, which translates into a reduction of the jet velocity and an increase in the jet size. The increase in the viscosity stabilizes the jet ejection, allowing a significant decrease in the flow rate, as occurs with glycerine. Specifically, $Q_{\textin{min}}=0.1$ ml/h and 0.025 ml/h for PVP 1\% and PVP 5\%, respectively. However, the minimum jet diameter for PVP 5\%, $d_j=5.9$ $\mu$m, is larger than that for PVP 1\%, $d_j=4.2$ $\mu$m. The latter is 2.7\% of the needle's inner diameter. This value can be further reduced by increasing the applied voltage. 

The dependence of the jet diameter on the flow rate significantly deviates from the scaling law $d_j\sim Q^{1/2}$ for the classical cone-jet mode of electrospray for Newtonian liquids \citep{G04a}. This implies that the jet velocity $v_j=4Q/(\pi d_j^2)$ depends on the injected flow rate. Specifically, $d_j\sim Q^{0.39}$ and $d_j\sim Q^{0.22}$ for the smaller flow rates of PVP 1\% and PVP 5\%, respectively (Fig.\ \ref{dj-Q}). This implies that $v_j\sim d_j^{0.62}$ for PVP 1\% and $v_j\sim d_j^{2.55}$ for PVP 5\%.

\citet{RVGM24} have recently analyzed the steady cone-jet mode of electrospray with ethyl cellulose dissolved in toluene using cylindrical capillaries with different diameters. The physical properties of this viscoelastic liquid are similar to those of PVP 1\%. Specifically, the zero-shear viscosities are in the order of tens of mPa, the extensional relaxation times are in the order of 1 ms, and the electrical conductivities are in the order of 1 $\mu$S/m. They conducted a systematic stability analysis and concluded that the minimum flow rate was achieved for $V\simeq 2$ kV and $H=1$ mm, similar to our experimental conditions. That minimum flow rate was around 2 ml/h, and the corresponding jet diameter was approximately 20 $\mu$m. The comparison with our results ($Q_{\textin{min}}=0.1$ ml/h and $d_j=4.2$ $\mu$m) suggests that using the beveled nozzle significantly reduces the minimum flow rate and jet diameter, increasing the printing spatial resolution. It must be noticed that the results of \citet{RVGM24} were obtained for a static substrate without the destabilizing effect of substrate displacement. Therefore, their analysis underestimated the minimum flow rate and jet diameter corresponding to the EHD jet printing. 

For the sake of comparison, we have conducted EHD jet printing experiments with PVP 5\% using a standard cylindrical capillary 160 $\mu$m in inner diameter for the experimental conditions considered in Fig.\ \ref{dj-Q}: $V=2$ kV and $H=1$ mm. All the experiments with flow rates below 0.5 ml/h suffered from the pull-out \citep{WHBMB12} and whipping \citep{MG20} instabilities. This flow rate is much larger than the minimum value $Q_{\textin{min}}=0.025$ ml/h for steady emission with the beveled nozzle.

\begin{figure}[hbt]
\begin{center}
\resizebox{0.4\textwidth}{!}{\includegraphics{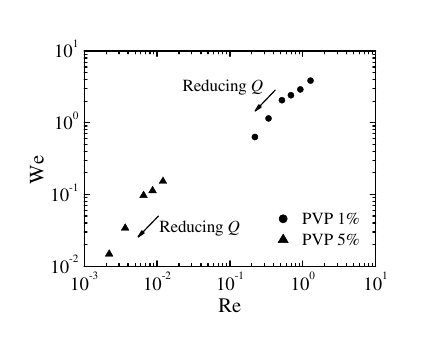}}
\end{center}
\caption{Weber number, We, and Reynolds number, Re, corresponding to all the experimental realizations.} 
\label{we}
\end{figure}

Due to the beveled nozzle stabilizing effect, the jet velocity can be decreased to values of the order of 0.1 m/s for PVP 5\%. Consequently, the jet inertia becomes negligible versus capillary and viscous stresses for the minimum flow rate. The PVP 1\% jet velocity can be decreased to values of the order of 1 m/s. In this case, inertia becomes comparable to surface tension and viscosity for $Q=Q_{\textin{min}}$. Figure \ref{we} shows the Weber number, We$=\rho v_j^2 d_j/\sigma$, and the Reynolds number, Re$=\rho v_j d_j/\mu$, corresponding to all the experimental realizations. As can be observed, We$\ll 1$ and Re$\ll 1$ for the minimum flow rates of PVP 5\%. We conclude that this viscoelastic solution is gently deposited on the substrate following a viscous capillary process.

\begin{figure}[hbt]
\begin{center}
\resizebox{0.45\textwidth}{!}{\includegraphics{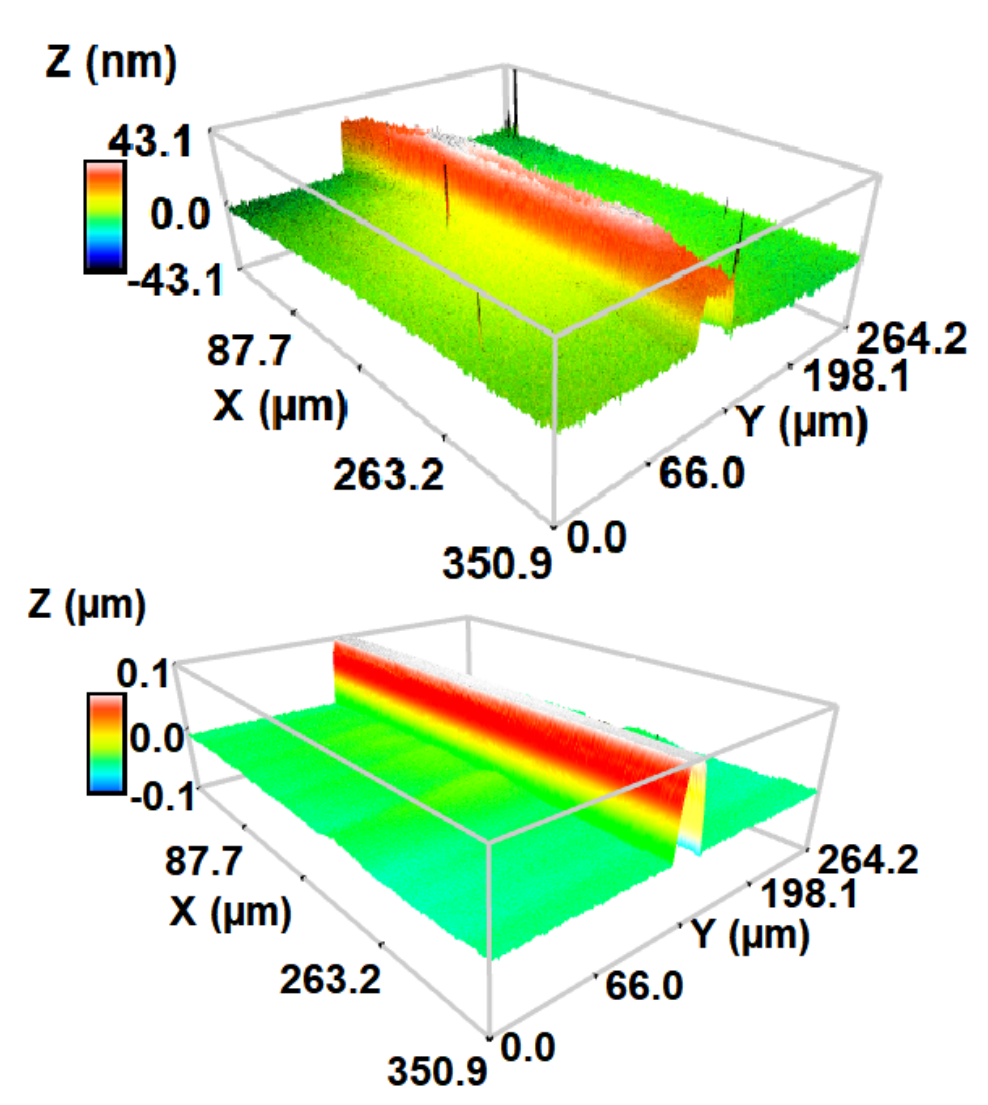}}
\end{center}
\caption{Optical perfilometry images of the printed lines for (PVP 1\%, $d_j=7.2$ $\mu$m) (upper image) and (PVP 5\%, $d_j=7.2$ $\mu$m) (lower image).} 
\label{3d}
\end{figure}

Now, we analyze the morphology of the printed lines for the three smaller flow rates of PVP 1\% and PVP 5\% (Fig.\ \ref{3d}). Lines with a high degree of uniformity were printed in all the cases. We determined the line width $w$ and height $h$ in different sections throughout the line. The standard deviations of these values were smaller than 5\%. The line height was decreased to tens of nanometers when PVP 1\% was printed. 

\begin{figure}[hbt]
\begin{center}
\resizebox{0.400\textwidth}{!}{\includegraphics{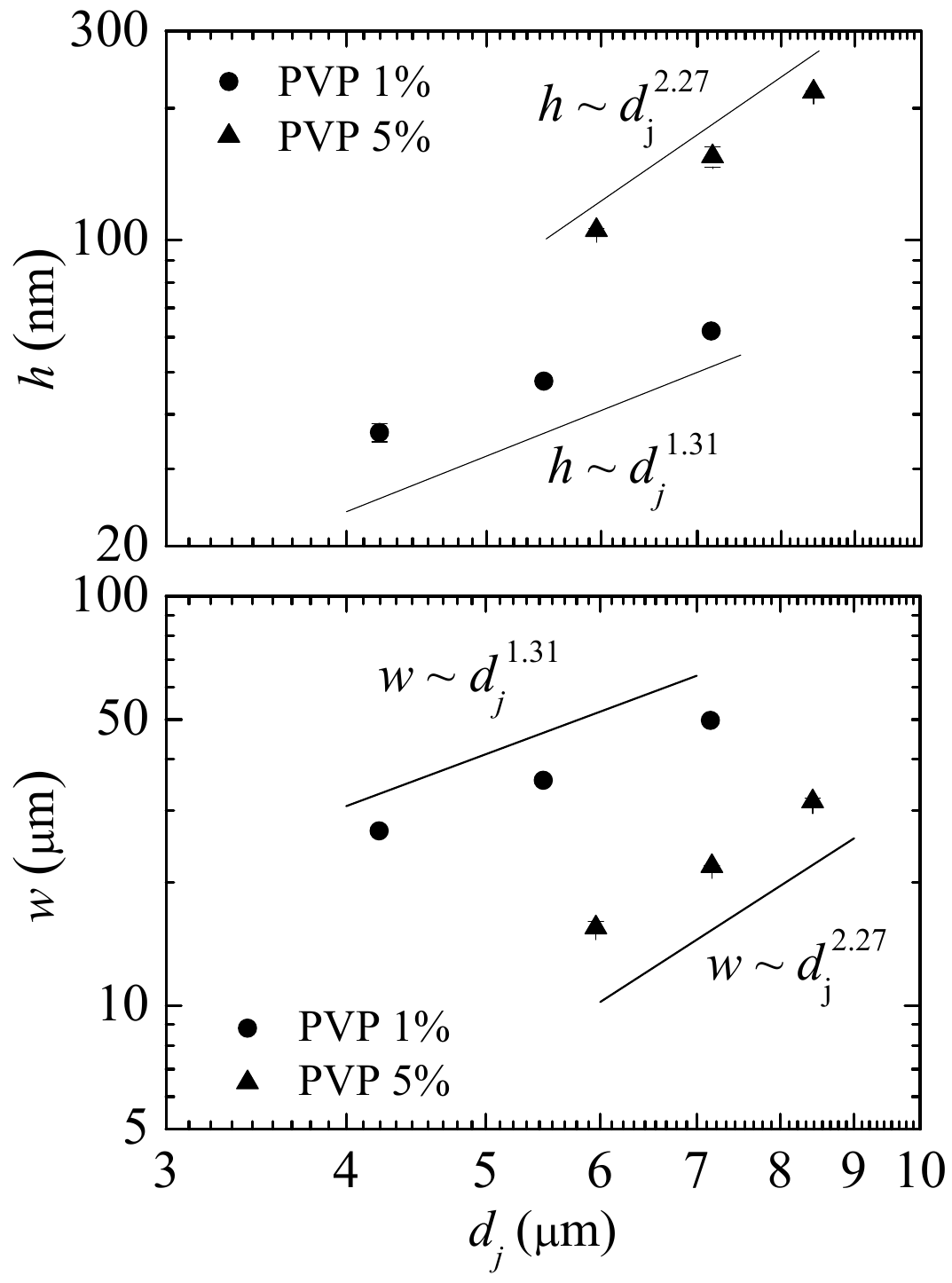}}
\end{center}
\caption{Height $h$ and width $w$ of the printed line as a function of the jet diameter $d_j$.} 
\label{h-w-dj}
\end{figure}

Interestingly, although smaller jet diameters were obtained with PVP 1\%, this did not translate into narrower lines (Fig.\ \ref{h-w-dj}). A larger amount of polymer in the solution caused the line to spread less on the substrate, facilitating vertical growth. In fact, the mean aspect ratio $h/w$ of PVP 1\% and PVP 5\% samples were 0.0013 and 0.0067, respectively. The narrowest line ($w\simeq 15.5$ $\mu$m) was printed with PVP 5\% and the lowest flow rate ($Q = 0.025$ ml/h). We conclude that the spatial resolution of EHD jet printing can be increased by increasing the polymer concentration and reducing the flow rate to its minimum value.

\begin{figure}[hbt]
\begin{center}
\resizebox{0.4\textwidth}{!}{\includegraphics{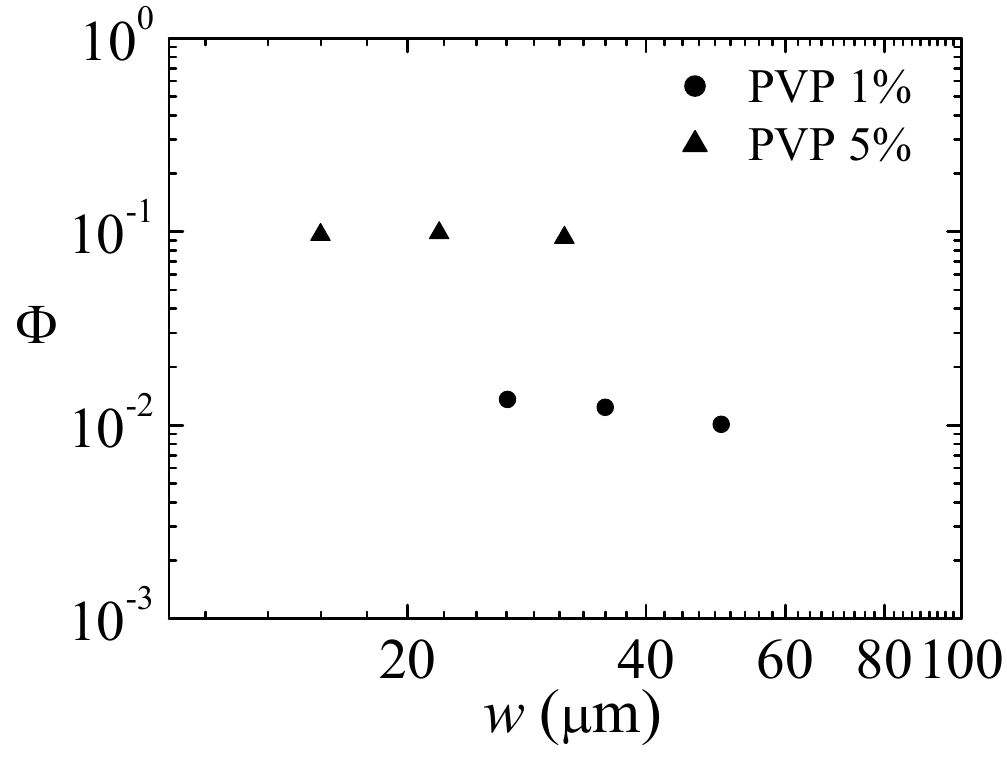}}
\end{center}
\caption{Volume fraction $\Phi$ remaining in printed line after evaporation as a function of the line width $w$.} 
\label{evap}
\end{figure}

The fraction of volume after evaporation, $\Phi$, can be calculated as $\Phi=A/A_0$, where $A$ and $A_0=Q/v_s$ are the line cross-section areas after and before evaporation, respectively. The fraction $\Phi$ does not significantly depend on the line width (Fig.\ \ref{evap}). It approximately coincides with the polymer mass fraction for PVP 1\% and significantly increases with the polymer concentration.

Consider the experiments for a given polymer concentration. All the variables affecting the line printing were kept constant except for the jet diameter $d_j$ and velocity $v_j$, which depended on the flow rate. The Weber number took very small values in all the experiments (Fig.\ \ref{we}). Thus, one expects the jet velocity $v_j$ to affect the line width $w$ only due to mass conservation: 
\begin{equation}
\label{con}
\pi d_j^2 v_j/4=(A/\Phi) v_s. 
\end{equation}

As observed in Fig.\ \ref{h-w-dj}, $w$ and $h$ exhibit a similar dependency on the jet diameter $d_j$ for a given polymer concentration, indicating that the aspect ratio $h/w$ is roughly constant. In other words, $h\sim w$ for a given polymer concentration, which suggests that $A\simeq k\; w^2$  with $k\simeq \text{const.}$ As shown in Fig.\ \ref{evap}, $\Phi\simeq \text{const.}$ in the analyzed experiments. Equation (\ref{con}) allows us to conclude that $w\sim d_j v_j^{1/2}$. As mentioned above, $v_j\sim d_j^{0.62}$ for PVP 1\% and $v_j\sim d_j^{2.55}$ for PVP 5\%. Therefore, $w\sim h\sim d_j^{1.31}$ for PVP 1\% and $w\sim h\sim d_j^{2.27}$ for PVP 5\%. These predictions agree with the experimental data shown in Fig.\ \ref{h-w-dj}.

\begin{figure*}[hbt]
\begin{center}
\resizebox{0.45\textwidth}{!}{\includegraphics{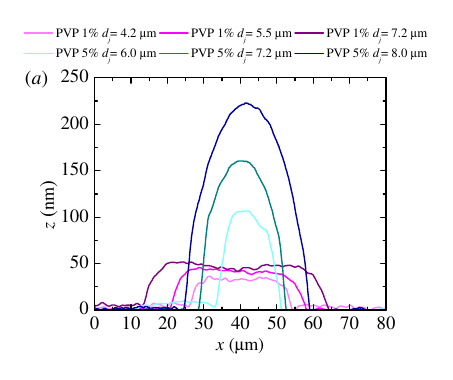}}\resizebox{0.46\textwidth}{!}{\includegraphics{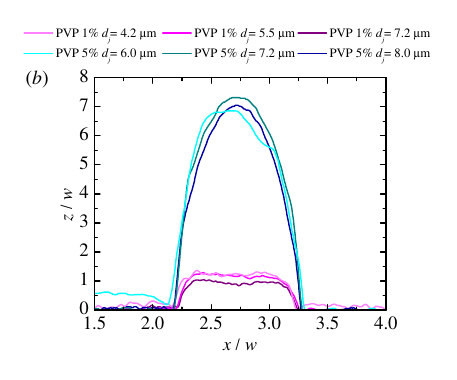}}
\end{center}
\caption{Printed line shape (a) and scaled printed line shape (b).} 
\label{shapes}
\end{figure*}

As mentioned above, the aspect ratio $h/w$ is roughly constant for a given polymer concentration. In fact, the printed line shapes are approximately similar. This can be appreciated in Fig.\ \ref{shapes}, where the shapes have been scaled with the corresponding width $w$. The contour of PVP 1\% obtained for the largest jet, $d_j=7.2$ $\mu$m, exhibits a depression in the line center, which is not present in the other two lines for the same concentration. This might be due to the significant effect of inertia on the PVP 1\% jet deposition (We$\sim 1$ and Re$\sim 1$).

\section{Conclusions}

We studied a dielectric beveled nozzle for electrospray and EHD jet printing. The electrospray experiments show that the beveled nozzle stabilizes the steady ejection mode in both the polar and viscous regimes. Our ejector considerably reduces the minimum flow rate. Specifically, the minimum flow rates are 17 and 125 times smaller than those of the standard electrospray configuration for octanol and glycerine, respectively. This leads to a significant increase in the aerosol charge-to-volume ratio, which has obvious implications in electrospray ionization. Jets with diameters around 1 $\mu$m were produced by injecting the liquid across a passage 160 times thicker, eliminating clogging effects.  

In the EHD jet printing experiments, viscoelastic threads a few microns in diameter were gently deposited on a moving hydrophilic substrate to print out lines tens of nanometers in thickness. Due to the stabilizing effect of the beveled nozzle, the minimum flow rate and jet diameter were much smaller than the respective values obtained with the cylindrical capillary. The flow rate of PVP 5\% could be decreased to the point at which deposition was fully dominated by capillarity/wetting and viscosity while inertia was negligible. This translated into a higher spatial resolution of the printing technique. The printed lines exhibited a high degree of uniformity. The experimental values of the line width and height were rationalized in terms of a simple scaling analysis. The printed line shapes obtained for the same polymer concentration and different flow rates were roughly similar. The comparison with previous results obtained using the cylindrical capillary \citep{RVGM24} suggests that the beveled nozzle significantly reduces the minimum flow rate and viscoelastic jet diameter, increasing the printing spatial resolution.

\blue{It must be pointed out that the present configuration of electrospray or EHD jet printing substantially differs from the classical one. In the classical configuration, the critical region where the liquid accelerates is much smaller than the size of the emitter. Therefore, a quasi-static electrified meniscus (the Taylor cone) held by surface tension connects the emitter and the jet. In the present configuration, the characteristic size of the emitter tip is commensurate with the jet, implying that the jet tapers directly from that tip. This eliminates the instability of the stagnation point located right in front of the cone-jet transition region of the classical configuration. Due to the no-slip boundary condition, the beveled emitter surface somehow replaces that stagnation point.}

The minimum flow rates achieved with the present configuration are much smaller than those predicted by the classical scaling laws. These deviations can be expected given the fundamental differences between the classical cone-jet mode of electrospray and the present configuration for $Q/Q_0\lesssim 0.1$, in which the Taylor cone is replaced with an extremely thin liquid film slipping over the beveled nozzle. Deriving the scaling laws \blue{to predict the minimum flow rate limit, the jet diameter, and the electric current} for the configuration proposed in this work is beyond the scope of the present paper.

There are several routes to reduce the size of low-conductivity jets electrosprayed with this configuration. One may optimize the design of the beveled nozzle, probably reducing the curvature radius of its tip. The nozzle wetting properties for a given electrosprayed liquid can also be optimized. We used an orifice of 200 $\mu$m in the counter-electrode to withdraw the ejected liquid. Ideally, the orifice diameter should decrease as the capillary-to-counter-electrode distance decreases to preserve similarity. Finally, we have not determined the optimal voltage and capillary-to-counter-electrode distance leading to the minimum flow rate (minimum jet diameter) in the stable regime. This optimal combination may lead to a further reduction of the jet diameter.

\vspace{1cm}

\newpage

\section*{CrediT authorship contribution statement}
\noindent D. Fern\'andez-Mart\'{\i}nez: Data curation; Formal analysis; Investigation; Methodology; Visualization; Writing - review \& editing.\\
E. J. Vega: Formal analysis; Investigation; Methodology; Visualization; Writing - review \& editing.\\
A. M. Ga\~n\'an-Calvo: Supervision; Writing - review \& editing. \\
J .M. Montanero: Conceptualization; Formal analysis; Investigation; Supervision; Writing - original draft; Writing - review \& editing.

\section*{Declaration of competing interest}
\noindent The authors declare that they have no known competing financial interests or personal relationships that could have appeared to influence the work reported in this paper.

\section*{Declaration of generative AI in scientific writing}
\noindent The authors declare that generative AI and AI-assisted technologies were not used in the writing process.

\section*{Data availability}
\noindent Data will be made available on request.

\section*{Acknowledgement}
This work was financially supported by the Spanish Ministry of Science, Innovation and Universities (grant no. PID2022-140951OB/AEI/10.13039/501100011033/FEDER, UE). 


\begin{thebibliography}{50}%
\makeatletter
\providecommand \@ifxundefined [1]{%
 \@ifx{#1\undefined}
}%
\providecommand \@ifnum [1]{%
 \ifnum #1\expandafter \@firstoftwo
 \else \expandafter \@secondoftwo
 \fi
}%
\providecommand \@ifx [1]{%
 \ifx #1\expandafter \@firstoftwo
 \else \expandafter \@secondoftwo
 \fi
}%
\providecommand \natexlab [1]{#1}%
\providecommand \enquote  [1]{``#1''}%
\providecommand \bibnamefont  [1]{#1}%
\providecommand \bibfnamefont [1]{#1}%
\providecommand \citenamefont [1]{#1}%
\providecommand \href@noop [0]{\@secondoftwo}%
\providecommand \href [0]{\begingroup \@sanitize@url \@href}%
\providecommand \@href[1]{\@@startlink{#1}\@@href}%
\providecommand \@@href[1]{\endgroup#1\@@endlink}%
\providecommand \@sanitize@url [0]{\catcode `\\12\catcode `\$12\catcode `\&12\catcode `\#12\catcode `\^12\catcode `\_12\catcode `\%12\relax}%
\providecommand \@@startlink[1]{}%
\providecommand \@@endlink[0]{}%
\providecommand \url  [0]{\begingroup\@sanitize@url \@url }%
\providecommand \@url [1]{\endgroup\@href {#1}{\urlprefix }}%
\providecommand \urlprefix  [0]{URL }%
\providecommand \Eprint [0]{\href }%
\providecommand \doibase [0]{http://dx.doi.org/}%
\providecommand \selectlanguage [0]{\@gobble}%
\providecommand \bibinfo  [0]{\@secondoftwo}%
\providecommand \bibfield  [0]{\@secondoftwo}%
\providecommand \translation [1]{[#1]}%
\providecommand \BibitemOpen [0]{}%
\providecommand \bibitemStop [0]{}%
\providecommand \bibitemNoStop [0]{.\EOS\space}%
\providecommand \EOS [0]{\spacefactor3000\relax}%
\providecommand \BibitemShut  [1]{\csname bibitem#1\endcsname}%
\let\auto@bib@innerbib\@empty
\bibitem [{\citenamefont {Ga{\~n}\'an-Calvo}\ \emph {et~al.}(2018)\citenamefont {Ga{\~n}\'an-Calvo}, \citenamefont {L\'opez-Herrera}, \citenamefont {Herrada}, \citenamefont {Ramos},\ and\ \citenamefont {Montanero}}]{GLHRM18}%
  \BibitemOpen
  \bibfield  {author} {\bibinfo {author} {\bibfnamefont {A.~M.}\ \bibnamefont {Ga{\~n}\'an-Calvo}}, \bibinfo {author} {\bibfnamefont {J.~M.}\ \bibnamefont {L\'opez-Herrera}}, \bibinfo {author} {\bibfnamefont {M.~A.}\ \bibnamefont {Herrada}}, \bibinfo {author} {\bibfnamefont {A.}~\bibnamefont {Ramos}}, \ and\ \bibinfo {author} {\bibfnamefont {J.~M.}\ \bibnamefont {Montanero}},\ }\bibfield  {title} {\enquote {\bibinfo {title} {Review on the physics of electrospray: from electrokinetics to the operating conditions of single and coaxial {Taylor} cone-jets, and {AC} electrospray},}\ }\href@noop {} {\bibfield  {journal} {\bibinfo  {journal} {J. Aerosol Sci.}\ }\textbf {\bibinfo {volume} {125}},\ \bibinfo {pages} {32--56} (\bibinfo {year} {2018})}\BibitemShut {NoStop}%
\bibitem [{\citenamefont {Rosell-Llompart}\ \emph {et~al.}(2018)\citenamefont {Rosell-Llompart}, \citenamefont {Grifoll},\ and\ \citenamefont {Loscertales}}]{RGL18}%
  \BibitemOpen
  \bibfield  {author} {\bibinfo {author} {\bibfnamefont {J.}~\bibnamefont {Rosell-Llompart}}, \bibinfo {author} {\bibfnamefont {J.}~\bibnamefont {Grifoll}}, \ and\ \bibinfo {author} {\bibfnamefont {I.~G.}\ \bibnamefont {Loscertales}},\ }\bibfield  {title} {\enquote {\bibinfo {title} {Electrosprays in the cone-jet mode: from {Taylor} cone formation to spray development},}\ }\href@noop {} {\bibfield  {journal} {\bibinfo  {journal} {J. Aerosol Sci.}\ } (\bibinfo {year} {2018})}\BibitemShut {NoStop}%
\bibitem [{\citenamefont {Ga{\~n}{\'a}n-Calvo}(2004)}]{G04a}%
  \BibitemOpen
  \bibfield  {author} {\bibinfo {author} {\bibfnamefont {A.~M.}\ \bibnamefont {Ga{\~n}{\'a}n-Calvo}},\ }\bibfield  {title} {\enquote {\bibinfo {title} {On the general scaling theory for electrospraying},}\ }\href@noop {} {\bibfield  {journal} {\bibinfo  {journal} {J. Fluid Mech.}\ }\textbf {\bibinfo {volume} {507}},\ \bibinfo {pages} {203--212} (\bibinfo {year} {2004})}\BibitemShut {NoStop}%
\bibitem [{\citenamefont {Ga{\~n}\'an-Calvo}\ \emph {et~al.}(2013)\citenamefont {Ga{\~n}\'an-Calvo}, \citenamefont {Rebollo-Mu{\~n}oz},\ and\ \citenamefont {Montanero}}]{GRM13}%
  \BibitemOpen
  \bibfield  {author} {\bibinfo {author} {\bibfnamefont {A.~M.}\ \bibnamefont {Ga{\~n}\'an-Calvo}}, \bibinfo {author} {\bibfnamefont {N.}~\bibnamefont {Rebollo-Mu{\~n}oz}}, \ and\ \bibinfo {author} {\bibfnamefont {J.~M.}\ \bibnamefont {Montanero}},\ }\bibfield  {title} {\enquote {\bibinfo {title} {Physical symmetries and scaling laws for the minimum or natural rate of flow and droplet size ejected by {Taylor} cone-jets},}\ }\href@noop {} {\bibfield  {journal} {\bibinfo  {journal} {New J. Phys.}\ }\textbf {\bibinfo {volume} {15}},\ \bibinfo {pages} {033035} (\bibinfo {year} {2013})}\BibitemShut {NoStop}%
\bibitem [{\citenamefont {Lozano}\ and\ \citenamefont {Mart\'{\i}nez-S\'anchez}(2005)}]{LM05}%
  \BibitemOpen
  \bibfield  {author} {\bibinfo {author} {\bibfnamefont {P.}~\bibnamefont {Lozano}}\ and\ \bibinfo {author} {\bibfnamefont {M.}~\bibnamefont {Mart\'{\i}nez-S\'anchez}},\ }\bibfield  {title} {\enquote {\bibinfo {title} {On the dynamic response of externally wetted ionic liquid ion sources},}\ }\href@noop {} {\bibfield  {journal} {\bibinfo  {journal} {J. Phys. D}\ }\textbf {\bibinfo {volume} {38}},\ \bibinfo {pages} {2371--2377} (\bibinfo {year} {2005})}\BibitemShut {NoStop}%
\bibitem [{\citenamefont {Lozano}\ and\ \citenamefont {Mart\'{\i}nez-S\'anchez}(2004)}]{LM04}%
  \BibitemOpen
  \bibfield  {author} {\bibinfo {author} {\bibfnamefont {P.}~\bibnamefont {Lozano}}\ and\ \bibinfo {author} {\bibfnamefont {M.}~\bibnamefont {Mart\'{\i}nez-S\'anchez}},\ }\bibfield  {title} {\enquote {\bibinfo {title} {Ionic liquid ion sources: Suppression of electrochemical reactions using voltage alternation},}\ }\href@noop {} {\bibfield  {journal} {\bibinfo  {journal} {J. Colloid Interface Sci.}\ }\textbf {\bibinfo {volume} {280}},\ \bibinfo {pages} {149--154} (\bibinfo {year} {2004})}\BibitemShut {NoStop}%
\bibitem [{\citenamefont {Castro}\ and\ \citenamefont {Mora}(2009)}]{CF09}%
  \BibitemOpen
  \bibfield  {author} {\bibinfo {author} {\bibfnamefont {S.}~\bibnamefont {Castro}}\ and\ \bibinfo {author} {\bibfnamefont {J.~Fern\'andez De~La}\ \bibnamefont {Mora}},\ }\bibfield  {title} {\enquote {\bibinfo {title} {Effect of tip curvature on ionic emissions from taylor cones of ionic liquids from externally wetted tungsten tips},}\ }\href@noop {} {\bibfield  {journal} {\bibinfo  {journal} {J. Appl. Phys.}\ }\textbf {\bibinfo {volume} {105}},\ \bibinfo {pages} {149--154} (\bibinfo {year} {2009})}\BibitemShut {NoStop}%
\bibitem [{\citenamefont {Vel\'{a}squez-Garc\'{i}a}\ \emph {et~al.}(2006)\citenamefont {Vel\'{a}squez-Garc\'{i}a}, \citenamefont {Akinwande},\ and\ \citenamefont {Mart\'{i}nez-S\'{a}nchez}}]{VAM06}%
  \BibitemOpen
  \bibfield  {author} {\bibinfo {author} {\bibfnamefont {L.~F.}\ \bibnamefont {Vel\'{a}squez-Garc\'{i}a}}, \bibinfo {author} {\bibfnamefont {A.~I.}\ \bibnamefont {Akinwande}}, \ and\ \bibinfo {author} {\bibfnamefont {M.}~\bibnamefont {Mart\'{i}nez-S\'{a}nchez}},\ }\bibfield  {title} {\enquote {\bibinfo {title} {A planar array of micro-fabricated electrospray emitters for thruster applications},}\ }\href@noop {} {\bibfield  {journal} {\bibinfo  {journal} {J. Microelectromech. Syst.}\ }\textbf {\bibinfo {volume} {15}},\ \bibinfo {pages} {1272--1280} (\bibinfo {year} {2006})}\BibitemShut {NoStop}%
\bibitem [{\citenamefont {Nakagawa}\ \emph {et~al.}(2017)\citenamefont {Nakagawa}, \citenamefont {Tsuchiya},\ and\ \citenamefont {Takao}}]{NTT17}%
  \BibitemOpen
  \bibfield  {author} {\bibinfo {author} {\bibfnamefont {K.}~\bibnamefont {Nakagawa}}, \bibinfo {author} {\bibfnamefont {T.}~\bibnamefont {Tsuchiya}}, \ and\ \bibinfo {author} {\bibfnamefont {Y.}~\bibnamefont {Takao}},\ }\bibfield  {title} {\enquote {\bibinfo {title} {Microfabricated emitter array for an ionic liquid electrospray thruster},}\ }\href@noop {} {\bibfield  {journal} {\bibinfo  {journal} {Jpn. J. Appl. Phys.}\ }\textbf {\bibinfo {volume} {56}},\ \bibinfo {pages} {149--154} (\bibinfo {year} {2017})}\BibitemShut {NoStop}%
\bibitem [{\citenamefont {Legge}\ and\ \citenamefont {Lozano}(2011)}]{LL11}%
  \BibitemOpen
  \bibfield  {author} {\bibinfo {author} {\bibfnamefont {R.}~\bibnamefont {Legge}}\ and\ \bibinfo {author} {\bibfnamefont {P.}~\bibnamefont {Lozano}},\ }\bibfield  {title} {\enquote {\bibinfo {title} {Electrospray propulsion based on emitters microfabricated in porous metals},}\ }\href@noop {} {\bibfield  {journal} {\bibinfo  {journal} {J. Propuls. Power}\ }\textbf {\bibinfo {volume} {27}},\ \bibinfo {pages} {485--495} (\bibinfo {year} {2011})}\BibitemShut {NoStop}%
\bibitem [{\citenamefont {Huang}\ \emph {et~al.}(2022)\citenamefont {Huang}, \citenamefont {Li}, \citenamefont {Li}, \citenamefont {Si}, \citenamefont {Xiong},\ and\ \citenamefont {Fan}}]{HLLSXF22}%
  \BibitemOpen
  \bibfield  {author} {\bibinfo {author} {\bibfnamefont {C.}~\bibnamefont {Huang}}, \bibinfo {author} {\bibfnamefont {J.}~\bibnamefont {Li}}, \bibinfo {author} {\bibfnamefont {M.}~\bibnamefont {Li}}, \bibinfo {author} {\bibfnamefont {T.}~\bibnamefont {Si}}, \bibinfo {author} {\bibfnamefont {C.}~\bibnamefont {Xiong}}, \ and\ \bibinfo {author} {\bibfnamefont {W.}~\bibnamefont {Fan}},\ }\bibfield  {title} {\enquote {\bibinfo {title} {Emission performance of ionic liquid electrospray thruster for micropropulsion},}\ }\href@noop {} {\bibfield  {journal} {\bibinfo  {journal} {J. Propuls. Power}\ }\textbf {\bibinfo {volume} {38}},\ \bibinfo {pages} {212--220} (\bibinfo {year} {2022})}\BibitemShut {NoStop}%
\bibitem [{\citenamefont {Courtney}\ and\ \citenamefont {Shea}(2015)}]{CS15}%
  \BibitemOpen
  \bibfield  {author} {\bibinfo {author} {\bibfnamefont {D.}~\bibnamefont {Courtney}}\ and\ \bibinfo {author} {\bibfnamefont {H.}~\bibnamefont {Shea}},\ }\bibfield  {title} {\enquote {\bibinfo {title} {Influences of porous reservoir laplace pressure on emissions from passively fed ionic liquid electrospray sources},}\ }\href@noop {} {\bibfield  {journal} {\bibinfo  {journal} {Appl. Phys. Lett.}\ }\textbf {\bibinfo {volume} {107}} (\bibinfo {year} {2015})}\BibitemShut {NoStop}%
\bibitem [{\citenamefont {Fenn}\ \emph {et~al.}(1989)\citenamefont {Fenn}, \citenamefont {Mann}, \citenamefont {Meng}, \citenamefont {Wong},\ and\ \citenamefont {Whitehouse}}]{FMMWW89}%
  \BibitemOpen
  \bibfield  {author} {\bibinfo {author} {\bibfnamefont {J.~B.}\ \bibnamefont {Fenn}}, \bibinfo {author} {\bibfnamefont {M.}~\bibnamefont {Mann}}, \bibinfo {author} {\bibfnamefont {C.~K.}\ \bibnamefont {Meng}}, \bibinfo {author} {\bibfnamefont {S.~F.}\ \bibnamefont {Wong}}, \ and\ \bibinfo {author} {\bibfnamefont {C.~M.}\ \bibnamefont {Whitehouse}},\ }\bibfield  {title} {\enquote {\bibinfo {title} {Electrospray ionization for mass spectrometry of large biomolecules},}\ }\href@noop {} {\bibfield  {journal} {\bibinfo  {journal} {Science}\ }\textbf {\bibinfo {volume} {246}},\ \bibinfo {pages} {64--71} (\bibinfo {year} {1989})}\BibitemShut {NoStop}%
\bibitem [{\citenamefont {Liu}\ \emph {et~al.}(2004)\citenamefont {Liu}, \citenamefont {Ro}, \citenamefont {Busman},\ and\ \citenamefont {Knapp}}]{LRBK04}%
  \BibitemOpen
  \bibfield  {author} {\bibinfo {author} {\bibfnamefont {J.}~\bibnamefont {Liu}}, \bibinfo {author} {\bibfnamefont {K.~W.}\ \bibnamefont {Ro}}, \bibinfo {author} {\bibfnamefont {M.}~\bibnamefont {Busman}}, \ and\ \bibinfo {author} {\bibfnamefont {D.~R.}\ \bibnamefont {Knapp}},\ }\bibfield  {title} {\enquote {\bibinfo {title} {Electrospray ionization with a pointed carbon fiber emitter},}\ }\href@noop {} {\bibfield  {journal} {\bibinfo  {journal} {Anal. Chem.}\ }\textbf {\bibinfo {volume} {76}},\ \bibinfo {pages} {3599--3606} (\bibinfo {year} {2004})}\BibitemShut {NoStop}%
\bibitem [{\citenamefont {Sen}\ \emph {et~al.}(2006)\citenamefont {Sen}, \citenamefont {Darabi}, \citenamefont {Knapp},\ and\ \citenamefont {Liu}}]{SDKL06}%
  \BibitemOpen
  \bibfield  {author} {\bibinfo {author} {\bibfnamefont {A.~K.}\ \bibnamefont {Sen}}, \bibinfo {author} {\bibfnamefont {J.}~\bibnamefont {Darabi}}, \bibinfo {author} {\bibfnamefont {D.~R.}\ \bibnamefont {Knapp}}, \ and\ \bibinfo {author} {\bibfnamefont {J.}~\bibnamefont {Liu}},\ }\bibfield  {title} {\enquote {\bibinfo {title} {Modeling and characterization of a carbon fiber emitter for electrospray ionization},}\ }\href@noop {} {\bibfield  {journal} {\bibinfo  {journal} {J. Micromech. Microeng.}\ }\textbf {\bibinfo {volume} {16}},\ \bibinfo {pages} {620--630} (\bibinfo {year} {2006})}\BibitemShut {NoStop}%
\bibitem [{\citenamefont {Schmidt}\ \emph {et~al.}(2003)\citenamefont {Schmidt}, \citenamefont {Karas},\ and\ \citenamefont {Dulcks}}]{SKD03}%
  \BibitemOpen
  \bibfield  {author} {\bibinfo {author} {\bibfnamefont {A.}~\bibnamefont {Schmidt}}, \bibinfo {author} {\bibfnamefont {M.}~\bibnamefont {Karas}}, \ and\ \bibinfo {author} {\bibfnamefont {T.}~\bibnamefont {Dulcks}},\ }\bibfield  {title} {\enquote {\bibinfo {title} {Effect of different solution flow rates on analyte ion signals in {Nano-ESI MS}, or: When does {ESI} turn into {Nano-ESI}?}}\ }\href@noop {} {\bibfield  {journal} {\bibinfo  {journal} {J. Am. Soc. Mass Spectrom.}\ }\textbf {\bibinfo {volume} {14}},\ \bibinfo {pages} {492–500} (\bibinfo {year} {2003})}\BibitemShut {NoStop}%
\bibitem [{\citenamefont {Gibson}\ \emph {et~al.}(2009)\citenamefont {Gibson}, \citenamefont {Mugo},\ and\ \citenamefont {Oleschuk}}]{GMO09}%
  \BibitemOpen
  \bibfield  {author} {\bibinfo {author} {\bibfnamefont {G.~T.~T.}\ \bibnamefont {Gibson}}, \bibinfo {author} {\bibfnamefont {S.~M.}\ \bibnamefont {Mugo}}, \ and\ \bibinfo {author} {\bibfnamefont {R.~D.}\ \bibnamefont {Oleschuk}},\ }\bibfield  {title} {\enquote {\bibinfo {title} {Nanoelectrospray emitters: Trends and perspectives},}\ }\href@noop {} {\bibfield  {journal} {\bibinfo  {journal} {Mass Spectrom. Rev.}\ }\textbf {\bibinfo {volume} {28}},\ \bibinfo {pages} {918–936} (\bibinfo {year} {2009})}\BibitemShut {NoStop}%
\bibitem [{\citenamefont {Sorensen}(1999)}]{S99}%
  \BibitemOpen
  \bibfield  {author} {\bibinfo {author} {\bibfnamefont {G.}~\bibnamefont {Sorensen}},\ }\bibfield  {title} {\enquote {\bibinfo {title} {Ion bombardment of electrosprayed coatings: an alternative to reactive sputtering?}}\ }\href@noop {} {\bibfield  {journal} {\bibinfo  {journal} {Surf. Coat. Tech.}\ }\textbf {\bibinfo {volume} {112}},\ \bibinfo {pages} {221--225} (\bibinfo {year} {1999})}\BibitemShut {NoStop}%
\bibitem [{\citenamefont {Rahman}\ \emph {et~al.}(2012)\citenamefont {Rahman}, \citenamefont {Khan}, \citenamefont {Muhammad}, \citenamefont {Jo},\ and\ \citenamefont {Choi}}]{RKMJC12}%
  \BibitemOpen
  \bibfield  {author} {\bibinfo {author} {\bibfnamefont {K.}~\bibnamefont {Rahman}}, \bibinfo {author} {\bibfnamefont {A.}~\bibnamefont {Khan}}, \bibinfo {author} {\bibfnamefont {N.~M.}\ \bibnamefont {Muhammad}}, \bibinfo {author} {\bibfnamefont {J.}~\bibnamefont {Jo}}, \ and\ \bibinfo {author} {\bibfnamefont {K.}~\bibnamefont {Choi}},\ }\bibfield  {title} {\enquote {\bibinfo {title} {Fine-resolution patterning of copper nanoparticles through electrohydrodynamic jet printing},}\ }\href@noop {} {\bibfield  {journal} {\bibinfo  {journal} {J. Micromech. Microeng.}\ }\textbf {\bibinfo {volume} {22}},\ \bibinfo {pages} {065012} (\bibinfo {year} {2012})}\BibitemShut {NoStop}%
\bibitem [{\citenamefont {Yang}\ \emph {et~al.}(2018)\citenamefont {Yang}, \citenamefont {Kim}, \citenamefont {Sajid}, \citenamefont {wan Kim}, \citenamefont {Aziz}, \citenamefont {Choi},\ and\ \citenamefont {Choi}}]{YKSKACC18}%
  \BibitemOpen
  \bibfield  {author} {\bibinfo {author} {\bibfnamefont {Y.~J.}\ \bibnamefont {Yang}}, \bibinfo {author} {\bibfnamefont {H.~C.}\ \bibnamefont {Kim}}, \bibinfo {author} {\bibfnamefont {M.}~\bibnamefont {Sajid}}, \bibinfo {author} {\bibfnamefont {S.}~\bibnamefont {wan Kim}}, \bibinfo {author} {\bibfnamefont {S.}~\bibnamefont {Aziz}}, \bibinfo {author} {\bibfnamefont {Y.~S.}\ \bibnamefont {Choi}}, \ and\ \bibinfo {author} {\bibfnamefont {K.~H.}\ \bibnamefont {Choi}},\ }\bibfield  {title} {\enquote {\bibinfo {title} {Drop-on-demand electrohydrodynamic printing of high resolution conductive micro patterns for mems repairing},}\ }\href@noop {} {\bibfield  {journal} {\bibinfo  {journal} {Int. J. Precis. Eng. Manuf.}\ }\textbf {\bibinfo {volume} {19}},\ \bibinfo {pages} {811--819} (\bibinfo {year} {2018})}\BibitemShut {NoStop}%
\bibitem [{\citenamefont {Li}\ \emph {et~al.}(2018)\citenamefont {Li}, \citenamefont {Jeong}, \citenamefont {Jang}, \citenamefont {Lim},\ and\ \citenamefont {Kim}}]{LJJLK18}%
  \BibitemOpen
  \bibfield  {author} {\bibinfo {author} {\bibfnamefont {X.}~\bibnamefont {Li}}, \bibinfo {author} {\bibfnamefont {Y.~J.}\ \bibnamefont {Jeong}}, \bibinfo {author} {\bibfnamefont {J.}~\bibnamefont {Jang}}, \bibinfo {author} {\bibfnamefont {S.}~\bibnamefont {Lim}}, \ and\ \bibinfo {author} {\bibfnamefont {S.~H.}\ \bibnamefont {Kim}},\ }\bibfield  {title} {\enquote {\bibinfo {title} {The effect of surfactants on electrohydrodynamic jet printing and the performance of organic field-effect transistors},}\ }\href@noop {} {\bibfield  {journal} {\bibinfo  {journal} {Phys. Chem. Chem. Phys.}\ }\textbf {\bibinfo {volume} {20}},\ \bibinfo {pages} {1210--1220} (\bibinfo {year} {2018})}\BibitemShut {NoStop}%
\bibitem [{\citenamefont {Can}\ \emph {et~al.}(2021)\citenamefont {Can}, \citenamefont {Kwack},\ and\ \citenamefont {Choi}}]{CKC21}%
  \BibitemOpen
  \bibfield  {author} {\bibinfo {author} {\bibfnamefont {T.~T.~T.}\ \bibnamefont {Can}}, \bibinfo {author} {\bibfnamefont {Y.}~\bibnamefont {Kwack}}, \ and\ \bibinfo {author} {\bibfnamefont {W.}~\bibnamefont {Choi}},\ }\bibfield  {title} {\enquote {\bibinfo {title} {Drop-on-demand patterning of mos2 using electrohydrodynamic jet printing for thin-film transistors},}\ }\href@noop {} {\bibfield  {journal} {\bibinfo  {journal} {Mater. Des.}\ }\textbf {\bibinfo {volume} {199}},\ \bibinfo {pages} {109408} (\bibinfo {year} {2021})}\BibitemShut {NoStop}%
\bibitem [{\citenamefont {Hong}\ \emph {et~al.}(2022)\citenamefont {Hong}, \citenamefont {Baek}, \citenamefont {Can}, \citenamefont {Choi},\ and\ \citenamefont {Kim}}]{HBCCK22}%
  \BibitemOpen
  \bibfield  {author} {\bibinfo {author} {\bibfnamefont {S.}~\bibnamefont {Hong}}, \bibinfo {author} {\bibfnamefont {S.}~\bibnamefont {Baek}}, \bibinfo {author} {\bibfnamefont {T.~T.~T.}\ \bibnamefont {Can}}, \bibinfo {author} {\bibfnamefont {W.-S.}\ \bibnamefont {Choi}}, \ and\ \bibinfo {author} {\bibfnamefont {S.}~\bibnamefont {Kim}},\ }\bibfield  {title} {\enquote {\bibinfo {title} {Fabrication of highly photosensitive mos2 photodetector films using rapid electrohydrodynamic-jet printing process},}\ }\href@noop {} {\bibfield  {journal} {\bibinfo  {journal} {Adv. Electron. Mater.}\ }\textbf {\bibinfo {volume} {8}},\ \bibinfo {pages} {2101063} (\bibinfo {year} {2022})}\BibitemShut {NoStop}%
\bibitem [{\citenamefont {Bi}\ \emph {et~al.}(2023)\citenamefont {Bi}, \citenamefont {Wang}, \citenamefont {Han}, \citenamefont {Wang}, \citenamefont {Tan}, \citenamefont {Shi}, \citenamefont {Jiang}, \citenamefont {He},\ and\ \citenamefont {Asare-Yeboah}}]{BWHWTSJHY23}%
  \BibitemOpen
  \bibfield  {author} {\bibinfo {author} {\bibfnamefont {S.}~\bibnamefont {Bi}}, \bibinfo {author} {\bibfnamefont {R.}~\bibnamefont {Wang}}, \bibinfo {author} {\bibfnamefont {X.}~\bibnamefont {Han}}, \bibinfo {author} {\bibfnamefont {Y.}~\bibnamefont {Wang}}, \bibinfo {author} {\bibfnamefont {D.}~\bibnamefont {Tan}}, \bibinfo {author} {\bibfnamefont {B.}~\bibnamefont {Shi}}, \bibinfo {author} {\bibfnamefont {C.}~\bibnamefont {Jiang}}, \bibinfo {author} {\bibfnamefont {Z.}~\bibnamefont {He}}, \ and\ \bibinfo {author} {\bibfnamefont {K.}~\bibnamefont {Asare-Yeboah}},\ }\bibfield  {title} {\enquote {\bibinfo {title} {Recent progress in electrohydrodynamic jet printing for printed electronics: From 0d to 3d materials},}\ }\href@noop {} {\bibfield  {journal} {\bibinfo  {journal} {Coatings}\ }\textbf {\bibinfo {volume} {13}},\ \bibinfo {pages} {1150} (\bibinfo {year} {2023})}\BibitemShut {NoStop}%
\bibitem [{\citenamefont {Hassan}\ \emph {et~al.}(2024)\citenamefont {Hassan}, \citenamefont {Sharipov},\ and\ \citenamefont {Ryu}}]{HSR24}%
  \BibitemOpen
  \bibfield  {author} {\bibinfo {author} {\bibfnamefont {R.~Ul}\ \bibnamefont {Hassan}}, \bibinfo {author} {\bibfnamefont {M.}~\bibnamefont {Sharipov}}, \ and\ \bibinfo {author} {\bibfnamefont {W.}~\bibnamefont {Ryu}},\ }\bibfield  {title} {\enquote {\bibinfo {title} {Electrohydrodynamic (ehd) printing of nanomaterial composite inks and their applications},}\ }\href@noop {} {\bibfield  {journal} {\bibinfo  {journal} {Micro and Nano Syst. Lett.}\ }\textbf {\bibinfo {volume} {11}},\ \bibinfo {pages} {2101589} (\bibinfo {year} {2024})}\BibitemShut {NoStop}%
\bibitem [{\citenamefont {Ringeisen}\ \emph {et~al.}(2006)\citenamefont {Ringeisen}, \citenamefont {Othon}, \citenamefont {Barron}, \citenamefont {Young},\ and\ \citenamefont {Spargo}}]{ROBYS06}%
  \BibitemOpen
  \bibfield  {author} {\bibinfo {author} {\bibfnamefont {B.~R.}\ \bibnamefont {Ringeisen}}, \bibinfo {author} {\bibfnamefont {C.~M.}\ \bibnamefont {Othon}}, \bibinfo {author} {\bibfnamefont {J.~A.}\ \bibnamefont {Barron}}, \bibinfo {author} {\bibfnamefont {D.}~\bibnamefont {Young}}, \ and\ \bibinfo {author} {\bibfnamefont {B.~J.}\ \bibnamefont {Spargo}},\ }\bibfield  {title} {\enquote {\bibinfo {title} {Jet-based methods to print living cells},}\ }\href@noop {} {\bibfield  {journal} {\bibinfo  {journal} {Biotechnol. J.}\ }\textbf {\bibinfo {volume} {1}},\ \bibinfo {pages} {930--948} (\bibinfo {year} {2006})}\BibitemShut {NoStop}%
\bibitem [{\citenamefont {Park}\ \emph {et~al.}(2007)\citenamefont {Park}, \citenamefont {Hardy}, \citenamefont {Kang}, \citenamefont {Barton}, \citenamefont {Adair}, \citenamefont {Mukhopadhyay}, \citenamefont {Lee}, \citenamefont {Strano}, \citenamefont {Alleyne}, \citenamefont {Georgiadis}, \citenamefont {Ferreira},\ and\ \citenamefont {Rogers}}]{Parketal}%
  \BibitemOpen
  \bibfield  {author} {\bibinfo {author} {\bibfnamefont {J.}~\bibnamefont {Park}}, \bibinfo {author} {\bibfnamefont {M.}~\bibnamefont {Hardy}}, \bibinfo {author} {\bibfnamefont {S.}~\bibnamefont {Kang}}, \bibinfo {author} {\bibfnamefont {K.}~\bibnamefont {Barton}}, \bibinfo {author} {\bibfnamefont {K.}~\bibnamefont {Adair}}, \bibinfo {author} {\bibfnamefont {D.}~\bibnamefont {Mukhopadhyay}}, \bibinfo {author} {\bibfnamefont {C.}~\bibnamefont {Lee}}, \bibinfo {author} {\bibfnamefont {M.~S.}\ \bibnamefont {Strano}}, \bibinfo {author} {\bibfnamefont {A.~G.}\ \bibnamefont {Alleyne}}, \bibinfo {author} {\bibfnamefont {J.~G.}\ \bibnamefont {Georgiadis}}, \bibinfo {author} {\bibfnamefont {P.~M.}\ \bibnamefont {Ferreira}}, \ and\ \bibinfo {author} {\bibfnamefont {J.~A.}\ \bibnamefont {Rogers}},\ }\bibfield  {title} {\enquote {\bibinfo {title} {High-resolution electrohydrodynamic jet printing},}\ }\href@noop {} {\bibfield  {journal} {\bibinfo  {journal} {Nat. Mater.}\ }\textbf {\bibinfo {volume} {6}},\ \bibinfo
  {pages} {782--789} (\bibinfo {year} {2007})}\BibitemShut {NoStop}%
\bibitem [{\citenamefont {Li}(2005)}]{L05}%
  \BibitemOpen
  \bibfield  {author} {\bibinfo {author} {\bibfnamefont {J.~L.}\ \bibnamefont {Li}},\ }\bibfield  {title} {\enquote {\bibinfo {title} {Formation and stabilization of an {EHD} jet from a nozzle with an inserted non-conductive fibre},}\ }\href@noop {} {\bibfield  {journal} {\bibinfo  {journal} {Aerosol Sci.}\ }\textbf {\bibinfo {volume} {36}},\ \bibinfo {pages} {373--386} (\bibinfo {year} {2005})}\BibitemShut {NoStop}%
\bibitem [{\citenamefont {Kim}\ \emph {et~al.}(2010)\citenamefont {Kim}, \citenamefont {Kim}, \citenamefont {Park},\ and\ \citenamefont {Hwang}}]{KKPH10}%
  \BibitemOpen
  \bibfield  {author} {\bibinfo {author} {\bibfnamefont {S.}~\bibnamefont {Kim}}, \bibinfo {author} {\bibfnamefont {Y.}~\bibnamefont {Kim}}, \bibinfo {author} {\bibfnamefont {J.}~\bibnamefont {Park}}, \ and\ \bibinfo {author} {\bibfnamefont {J.}~\bibnamefont {Hwang}},\ }\bibfield  {title} {\enquote {\bibinfo {title} {Design and evaluation of single nozzle with a non-conductive tip for reducing applied voltage and pattern width in electrohydrodynamic jet printing (ehdp)},}\ }\href@noop {} {\bibfield  {journal} {\bibinfo  {journal} {J. Micromech. Microeng.}\ }\textbf {\bibinfo {volume} {20}} (\bibinfo {year} {2010})}\BibitemShut {NoStop}%
\bibitem [{\citenamefont {Wang}\ \emph {et~al.}(2012)\citenamefont {Wang}, \citenamefont {Qiu}, \citenamefont {Pei}, \citenamefont {Su}, \citenamefont {Zhan}, \citenamefont {Lv},\ and\ \citenamefont {Sun}}]{WQPSZLS12}%
  \BibitemOpen
  \bibfield  {author} {\bibinfo {author} {\bibfnamefont {L.}~\bibnamefont {Wang}}, \bibinfo {author} {\bibfnamefont {Y.}~\bibnamefont {Qiu}}, \bibinfo {author} {\bibfnamefont {Y.}~\bibnamefont {Pei}}, \bibinfo {author} {\bibfnamefont {Y.}~\bibnamefont {Su}}, \bibinfo {author} {\bibfnamefont {Z.}~\bibnamefont {Zhan}}, \bibinfo {author} {\bibfnamefont {W.}~\bibnamefont {Lv}}, \ and\ \bibinfo {author} {\bibfnamefont {D.}~\bibnamefont {Sun}},\ }\bibfield  {title} {\enquote {\bibinfo {title} {A novel electrohydrodynamic printing jet head with retractable needle},}\ }\href@noop {} {\bibfield  {journal} {\bibinfo  {journal} {Proc. Inst. Mech. Eng., Part N}\ }\textbf {\bibinfo {volume} {225}},\ \bibinfo {pages} {85--88} (\bibinfo {year} {2012})}\BibitemShut {NoStop}%
\bibitem [{\citenamefont {Acero}\ \emph {et~al.}(2013)\citenamefont {Acero}, \citenamefont {Rebollo-Mu{\~n}oz}, \citenamefont {Montanero}, \citenamefont {Ga{\~n}{\'a}n-Calvo},\ and\ \citenamefont {Vega}}]{ARMGV13}%
  \BibitemOpen
  \bibfield  {author} {\bibinfo {author} {\bibfnamefont {A.~J.}\ \bibnamefont {Acero}}, \bibinfo {author} {\bibfnamefont {N.}~\bibnamefont {Rebollo-Mu{\~n}oz}}, \bibinfo {author} {\bibfnamefont {J.~M.}\ \bibnamefont {Montanero}}, \bibinfo {author} {\bibfnamefont {A.~M.}\ \bibnamefont {Ga{\~n}{\'a}n-Calvo}}, \ and\ \bibinfo {author} {\bibfnamefont {E.~J.}\ \bibnamefont {Vega}},\ }\bibfield  {title} {\enquote {\bibinfo {title} {A new flow focusing technique to produce very thin jets},}\ }\href@noop {} {\bibfield  {journal} {\bibinfo  {journal} {J. Micromech. Microeng.}\ }\textbf {\bibinfo {volume} {23}},\ \bibinfo {pages} {065009} (\bibinfo {year} {2013})}\BibitemShut {NoStop}%
\bibitem [{\citenamefont {Rebollo-Mu{\~n}oz}\ \emph {et~al.}(2016)\citenamefont {Rebollo-Mu{\~n}oz}, \citenamefont {Acero}, \citenamefont {{Marcos de Le\'on}}, \citenamefont {Montanero},\ and\ \citenamefont {Ga{\~n}\'an-Calvo}}]{RAMMG16}%
  \BibitemOpen
  \bibfield  {author} {\bibinfo {author} {\bibfnamefont {N.}~\bibnamefont {Rebollo-Mu{\~n}oz}}, \bibinfo {author} {\bibfnamefont {A.~J.}\ \bibnamefont {Acero}}, \bibinfo {author} {\bibfnamefont {J.~Z.}\ \bibnamefont {{Marcos de Le\'on}}}, \bibinfo {author} {\bibfnamefont {J.M.}\ \bibnamefont {Montanero}}, \ and\ \bibinfo {author} {\bibfnamefont {A.M.}\ \bibnamefont {Ga{\~n}\'an-Calvo}},\ }\bibfield  {title} {\enquote {\bibinfo {title} {A hybrid flow focusing nozzle design to produce micron and sub-micron capillary jets},}\ }\href@noop {} {\bibfield  {journal} {\bibinfo  {journal} {Int. J. Mass Spectrom.}\ }\textbf {\bibinfo {volume} {403}},\ \bibinfo {pages} {32--38} (\bibinfo {year} {2016})}\BibitemShut {NoStop}%
\bibitem [{\citenamefont {Nazari}\ \emph {et~al.}(2020)\citenamefont {Nazari}, \citenamefont {Zaare}, \citenamefont {Alvarez}, \citenamefont {Karpos}, \citenamefont {Engelman}, \citenamefont {Madsen}, \citenamefont {Nelson}, \citenamefont {Spence}, \citenamefont {Weierstall}, \citenamefont {Adrian},\ and\ \citenamefont {Kirian}}]{NZAKEMNSWAK20}%
  \BibitemOpen
  \bibfield  {author} {\bibinfo {author} {\bibfnamefont {R.}~\bibnamefont {Nazari}}, \bibinfo {author} {\bibfnamefont {S.}~\bibnamefont {Zaare}}, \bibinfo {author} {\bibfnamefont {R.~C.}\ \bibnamefont {Alvarez}}, \bibinfo {author} {\bibfnamefont {K.}~\bibnamefont {Karpos}}, \bibinfo {author} {\bibfnamefont {T.}~\bibnamefont {Engelman}}, \bibinfo {author} {\bibfnamefont {C.}~\bibnamefont {Madsen}}, \bibinfo {author} {\bibfnamefont {G.}~\bibnamefont {Nelson}}, \bibinfo {author} {\bibfnamefont {J.~C.~H.}\ \bibnamefont {Spence}}, \bibinfo {author} {\bibfnamefont {U.}~\bibnamefont {Weierstall}}, \bibinfo {author} {\bibfnamefont {R.~J.}\ \bibnamefont {Adrian}}, \ and\ \bibinfo {author} {\bibfnamefont {R.~A.}\ \bibnamefont {Kirian}},\ }\bibfield  {title} {\enquote {\bibinfo {title} {{3D} printing of gas-dynamic virtual nozzles and optical characterization of high-speed microjets},}\ }\href@noop {} {\bibfield  {journal} {\bibinfo  {journal} {Opt. Express}\ }\textbf {\bibinfo {volume} {28}},\ \bibinfo {pages}
  {21749--21765} (\bibinfo {year} {2020})}\BibitemShut {NoStop}%
\bibitem [{\citenamefont {Chen}\ \emph {et~al.}(1999)\citenamefont {Chen}, \citenamefont {Emond}, \citenamefont {Kelder},\ and\ \citenamefont {Schoonman}}]{CEKMS99}%
  \BibitemOpen
  \bibfield  {author} {\bibinfo {author} {\bibfnamefont {C.H.}\ \bibnamefont {Chen}}, \bibinfo {author} {\bibfnamefont {M.H.J.}\ \bibnamefont {Emond}}, \bibinfo {author} {\bibfnamefont {E.M.}\ \bibnamefont {Kelder}}, \ and\ \bibinfo {author} {\bibfnamefont {J.}~\bibnamefont {Schoonman}},\ }\bibfield  {title} {\enquote {\bibinfo {title} {Electrostatic sol–spray deposition of nanostructured ceramic thin films},}\ }\href@noop {} {\bibfield  {journal} {\bibinfo  {journal} {J. Aerosol. Sci.}\ }\textbf {\bibinfo {volume} {30}},\ \bibinfo {pages} {959--967} (\bibinfo {year} {1999})}\BibitemShut {NoStop}%
\bibitem [{\citenamefont {Ksapabutr}\ \emph {et~al.}(2008)\citenamefont {Ksapabutr}, \citenamefont {Panapoy}, \citenamefont {Choncharoen}, \citenamefont {Wongkasemjit},\ and\ \citenamefont {Traversa}}]{KPCWT08}%
  \BibitemOpen
  \bibfield  {author} {\bibinfo {author} {\bibfnamefont {B.}~\bibnamefont {Ksapabutr}}, \bibinfo {author} {\bibfnamefont {M.}~\bibnamefont {Panapoy}}, \bibinfo {author} {\bibfnamefont {K.}~\bibnamefont {Choncharoen}}, \bibinfo {author} {\bibfnamefont {S.}~\bibnamefont {Wongkasemjit}}, \ and\ \bibinfo {author} {\bibfnamefont {E.}~\bibnamefont {Traversa}},\ }\bibfield  {title} {\enquote {\bibinfo {title} {Investigation of nozzle shape effect on sm0.1ce0.9o1.95 thin film prepared by electrostatic spray deposition},}\ }\href@noop {} {\bibfield  {journal} {\bibinfo  {journal} {Thin Solid Films}\ }\textbf {\bibinfo {volume} {516}},\ \bibinfo {pages} {5618--5624} (\bibinfo {year} {2008})}\BibitemShut {NoStop}%
\bibitem [{\citenamefont {Leeuwenburgh}\ \emph {et~al.}(2003)\citenamefont {Leeuwenburgh}, \citenamefont {Wolke}, \citenamefont {Schoonman},\ and\ \citenamefont {Jansen}}]{LWSJ03}%
  \BibitemOpen
  \bibfield  {author} {\bibinfo {author} {\bibfnamefont {S.}~\bibnamefont {Leeuwenburgh}}, \bibinfo {author} {\bibfnamefont {J.}~\bibnamefont {Wolke}}, \bibinfo {author} {\bibfnamefont {J.}~\bibnamefont {Schoonman}}, \ and\ \bibinfo {author} {\bibfnamefont {J.}~\bibnamefont {Jansen}},\ }\bibfield  {title} {\enquote {\bibinfo {title} {Electrostatic spray deposition (esd) of calcium phosphate coatings},}\ }\href@noop {} {\bibfield  {journal} {\bibinfo  {journal} {J. Biomed. Mater. Res. A}\ }\textbf {\bibinfo {volume} {66}},\ \bibinfo {pages} {330--334} (\bibinfo {year} {2003})}\BibitemShut {NoStop}%
\bibitem [{\citenamefont {Jaworek}\ \emph {et~al.}(2018)\citenamefont {Jaworek}, \citenamefont {Sobczyk},\ and\ \citenamefont {Krupa}}]{JSK18}%
  \BibitemOpen
  \bibfield  {author} {\bibinfo {author} {\bibfnamefont {A.}~\bibnamefont {Jaworek}}, \bibinfo {author} {\bibfnamefont {A.~T.}\ \bibnamefont {Sobczyk}}, \ and\ \bibinfo {author} {\bibfnamefont {A.}~\bibnamefont {Krupa}},\ }\bibfield  {title} {\enquote {\bibinfo {title} {Electrospray application to powder production and surface coating},}\ }\href@noop {} {\bibfield  {journal} {\bibinfo  {journal} {J. Aerosol. Sci.}\ }\textbf {\bibinfo {volume} {125}},\ \bibinfo {pages} {57--92} (\bibinfo {year} {2018})}\BibitemShut {NoStop}%
\bibitem [{\citenamefont {Youn}\ \emph {et~al.}(2009)\citenamefont {Youn}, \citenamefont {Kim}, \citenamefont {Yang}, \citenamefont {Lim}, \citenamefont {Kim}, \citenamefont {Ahn}, \citenamefont {Sim}, \citenamefont {Ryu}, \citenamefont {Shin},\ and\ \citenamefont {Yoo}}]{YKYLKASRSY09}%
  \BibitemOpen
  \bibfield  {author} {\bibinfo {author} {\bibfnamefont {D.}~\bibnamefont {Youn}}, \bibinfo {author} {\bibfnamefont {S.}~\bibnamefont {Kim}}, \bibinfo {author} {\bibfnamefont {Y.}~\bibnamefont {Yang}}, \bibinfo {author} {\bibfnamefont {S.}~\bibnamefont {Lim}}, \bibinfo {author} {\bibfnamefont {S.}~\bibnamefont {Kim}}, \bibinfo {author} {\bibfnamefont {S.}~\bibnamefont {Ahn}}, \bibinfo {author} {\bibfnamefont {H.}~\bibnamefont {Sim}}, \bibinfo {author} {\bibfnamefont {S.}~\bibnamefont {Ryu}}, \bibinfo {author} {\bibfnamefont {D.}~\bibnamefont {Shin}}, \ and\ \bibinfo {author} {\bibfnamefont {J.}~\bibnamefont {Yoo}},\ }\bibfield  {title} {\enquote {\bibinfo {title} {Electrohydrodynamic micropatterning of silver ink using near-field electrohydrodynamic jet printing with tilted-outlet nozzle},}\ }\href@noop {} {\bibfield  {journal} {\bibinfo  {journal} {Appl. Phys. A-Mater.}\ }\textbf {\bibinfo {volume} {96}},\ \bibinfo {pages} {933--938} (\bibinfo {year} {2009})}\BibitemShut {NoStop}%
\bibitem [{\citenamefont {Kim}\ \emph {et~al.}(2015)\citenamefont {Kim}, \citenamefont {Jang},\ and\ \citenamefont {Oh}}]{KAO15}%
  \BibitemOpen
  \bibfield  {author} {\bibinfo {author} {\bibfnamefont {Y.}~\bibnamefont {Kim}}, \bibinfo {author} {\bibfnamefont {S.}~\bibnamefont {Jang}}, \ and\ \bibinfo {author} {\bibfnamefont {J.}~\bibnamefont {Oh}},\ }\bibfield  {title} {\enquote {\bibinfo {title} {High-resolution electrohydrodynamic printing of silver nanoparticle ink via commercial hypodermic needles},}\ }\href@noop {} {\bibfield  {journal} {\bibinfo  {journal} {Appl. Phys. Lett.}\ }\textbf {\bibinfo {volume} {106}} (\bibinfo {year} {2015})}\BibitemShut {NoStop}%
\bibitem [{\citenamefont {Rehmani}\ and\ \citenamefont {Arif}(2021)}]{RA21}%
  \BibitemOpen
  \bibfield  {author} {\bibinfo {author} {\bibfnamefont {M.}~\bibnamefont {Rehmani}}\ and\ \bibinfo {author} {\bibfnamefont {K.}~\bibnamefont {Arif}},\ }\bibfield  {title} {\enquote {\bibinfo {title} {High resolution electrohydrodynamic printing of conductive ink with an aligned aperture coaxial printhead},}\ }\href@noop {} {\bibfield  {journal} {\bibinfo  {journal} {J. Adv. Manuf. Technol.}\ }\textbf {\bibinfo {volume} {115}},\ \bibinfo {pages} {{2785--2800}} (\bibinfo {year} {2021})}\BibitemShut {NoStop}%
\bibitem [{\citenamefont {Vu}\ \emph {et~al.}(2022)\citenamefont {Vu}, \citenamefont {Nguyen}, \citenamefont {Fastier-Wooller}, \citenamefont {Tran}, \citenamefont {Nguyen}, \citenamefont {Nguyen}, \citenamefont {Nguyen}, \citenamefont {Nguyen}, \citenamefont {Dinh}, \citenamefont {Bui}, \citenamefont {Zhong}, \citenamefont {Phan}, \citenamefont {Nguyen}, \citenamefont {Dao},\ and\ \citenamefont {Dau}}]{Vuetal22}%
  \BibitemOpen
  \bibfield  {author} {\bibinfo {author} {\bibfnamefont {T.}~\bibnamefont {Vu}}, \bibinfo {author} {\bibfnamefont {H.}~\bibnamefont {Nguyen}}, \bibinfo {author} {\bibfnamefont {J.~W.}\ \bibnamefont {Fastier-Wooller}}, \bibinfo {author} {\bibfnamefont {C.}~\bibnamefont {Tran}}, \bibinfo {author} {\bibfnamefont {T.}~\bibnamefont {Nguyen}}, \bibinfo {author} {\bibfnamefont {H.}~\bibnamefont {Nguyen}}, \bibinfo {author} {\bibfnamefont {T.}~\bibnamefont {Nguyen}}, \bibinfo {author} {\bibfnamefont {T.}~\bibnamefont {Nguyen}}, \bibinfo {author} {\bibfnamefont {T.}~\bibnamefont {Dinh}}, \bibinfo {author} {\bibfnamefont {T.~T.}\ \bibnamefont {Bui}}, \bibinfo {author} {\bibfnamefont {Y.}~\bibnamefont {Zhong}}, \bibinfo {author} {\bibfnamefont {H.}~\bibnamefont {Phan}}, \bibinfo {author} {\bibfnamefont {N.}~\bibnamefont {Nguyen}}, \bibinfo {author} {\bibfnamefont {D.~V.}\ \bibnamefont {Dao}}, \ and\ \bibinfo {author} {\bibfnamefont {V.}~\bibnamefont {Dau}},\ }\bibfield  {title} {\enquote {\bibinfo {title} {Enhanced
  electrohydrodynamics for electrospinning a highly sensitive flexible fiber-based piezoelectric sensor},}\ }\href@noop {} {\bibfield  {journal} {\bibinfo  {journal} {ACS Appl. Electron. Mater.}\ }\textbf {\bibinfo {volume} {4}},\ \bibinfo {pages} {1301--1310} (\bibinfo {year} {2022})}\BibitemShut {NoStop}%
\bibitem [{\citenamefont {Rebollo-Muñoz}\ \emph {et~al.}(2013)\citenamefont {Rebollo-Muñoz}, \citenamefont {Montanero},\ and\ \citenamefont {Gañán-Calvo}}]{RMG13}%
  \BibitemOpen
  \bibfield  {author} {\bibinfo {author} {\bibfnamefont {N.}~\bibnamefont {Rebollo-Muñoz}}, \bibinfo {author} {\bibfnamefont {J.M.}\ \bibnamefont {Montanero}}, \ and\ \bibinfo {author} {\bibfnamefont {A.M.}\ \bibnamefont {Gañán-Calvo}},\ }\bibfield  {title} {\enquote {\bibinfo {title} {On the use of hypodermic needles in electrospray},}\ }\href@noop {} {\bibfield  {journal} {\bibinfo  {journal} {EPJ Web of Conferences}\ }\textbf {\bibinfo {volume} {45}} (\bibinfo {year} {2013})}\BibitemShut {NoStop}%
\bibitem [{\citenamefont {Cabezas}\ \emph {et~al.}(2004)\citenamefont {Cabezas}, \citenamefont {Bateni}, \citenamefont {Montanero},\ and\ \citenamefont {Neumann}}]{CBMN04}%
  \BibitemOpen
  \bibfield  {author} {\bibinfo {author} {\bibfnamefont {M.~G.}\ \bibnamefont {Cabezas}}, \bibinfo {author} {\bibfnamefont {A.}~\bibnamefont {Bateni}}, \bibinfo {author} {\bibfnamefont {J.~M.}\ \bibnamefont {Montanero}}, \ and\ \bibinfo {author} {\bibfnamefont {A.~W.}\ \bibnamefont {Neumann}},\ }\bibfield  {title} {\enquote {\bibinfo {title} {A new drop-shape methodology for surface tension measurement},}\ }\href@noop {} {\bibfield  {journal} {\bibinfo  {journal} {Appl. Surf. Sci.}\ }\textbf {\bibinfo {volume} {238}},\ \bibinfo {pages} {480--484} (\bibinfo {year} {2004})}\BibitemShut {NoStop}%
\bibitem [{\citenamefont {Sousa}\ \emph {et~al.}(2017)\citenamefont {Sousa}, \citenamefont {Vega}, \citenamefont {Sousa}, \citenamefont {Montanero},\ and\ \citenamefont {Alves}}]{SVSMA17}%
  \BibitemOpen
  \bibfield  {author} {\bibinfo {author} {\bibfnamefont {P.~C.}\ \bibnamefont {Sousa}}, \bibinfo {author} {\bibfnamefont {E.~J.}\ \bibnamefont {Vega}}, \bibinfo {author} {\bibfnamefont {R.~G.}\ \bibnamefont {Sousa}}, \bibinfo {author} {\bibfnamefont {J.~M.}\ \bibnamefont {Montanero}}, \ and\ \bibinfo {author} {\bibfnamefont {M.~A.}\ \bibnamefont {Alves}},\ }\bibfield  {title} {\enquote {\bibinfo {title} {Measurement of relaxation times in extensional flow of weakly viscoelastic polymer solutions},}\ }\href@noop {} {\bibfield  {journal} {\bibinfo  {journal} {Rheol. Acta}\ }\textbf {\bibinfo {volume} {56}},\ \bibinfo {pages} {11--20} (\bibinfo {year} {2017})}\BibitemShut {NoStop}%
\bibitem [{\citenamefont {Ponce-Torres}\ \emph {et~al.}(2018)\citenamefont {Ponce-Torres}, \citenamefont {Rebollo-Mu{\~n}oz}, \citenamefont {Herrada}, \citenamefont {Ga{\~n}\'an-Calvo},\ and\ \citenamefont {Montanero}}]{PRHGM18}%
  \BibitemOpen
  \bibfield  {author} {\bibinfo {author} {\bibfnamefont {A.}~\bibnamefont {Ponce-Torres}}, \bibinfo {author} {\bibfnamefont {N.}~\bibnamefont {Rebollo-Mu{\~n}oz}}, \bibinfo {author} {\bibfnamefont {M.~A.}\ \bibnamefont {Herrada}}, \bibinfo {author} {\bibfnamefont {A.~M.}\ \bibnamefont {Ga{\~n}\'an-Calvo}}, \ and\ \bibinfo {author} {\bibfnamefont {J.~M.}\ \bibnamefont {Montanero}},\ }\bibfield  {title} {\enquote {\bibinfo {title} {The steady cone-jet mode of electrospraying close to the minimum volume stability limit},}\ }\href@noop {} {\bibfield  {journal} {\bibinfo  {journal} {J. Fluid Mech.}\ }\textbf {\bibinfo {volume} {857}},\ \bibinfo {pages} {142--172} (\bibinfo {year} {2018})}\BibitemShut {NoStop}%
\bibitem [{\citenamefont {{Fernandez de la Mora}}\ and\ \citenamefont {Loscertales}(1994)}]{FL94}%
  \BibitemOpen
  \bibfield  {author} {\bibinfo {author} {\bibfnamefont {J.}~\bibnamefont {{Fernandez de la Mora}}}\ and\ \bibinfo {author} {\bibfnamefont {I.~G.}\ \bibnamefont {Loscertales}},\ }\bibfield  {title} {\enquote {\bibinfo {title} {The current transmitted through an electrified conical meniscus},}\ }\href@noop {} {\bibfield  {journal} {\bibinfo  {journal} {J . Fluid Mech.}\ }\textbf {\bibinfo {volume} {260}},\ \bibinfo {pages} {155--184} (\bibinfo {year} {1994})}\BibitemShut {NoStop}%
\bibitem [{\citenamefont {Higuera}(2017)}]{H17}%
  \BibitemOpen
  \bibfield  {author} {\bibinfo {author} {\bibfnamefont {F.~J.}\ \bibnamefont {Higuera}},\ }\bibfield  {title} {\enquote {\bibinfo {title} {Qualitative analysis of the minimum flow rate of a cone-jet of a very polar liquid},}\ }\href@noop {} {\bibfield  {journal} {\bibinfo  {journal} {J. Fluid Mech.}\ }\textbf {\bibinfo {volume} {816}},\ \bibinfo {pages} {428--441} (\bibinfo {year} {2017})}\BibitemShut {NoStop}%
\bibitem [{\citenamefont {Rijo}\ \emph {et~al.}(2024)\citenamefont {Rijo}, \citenamefont {Vega}, \citenamefont {Galindo-Rosales},\ and\ \citenamefont {Montanero}}]{RVGM24}%
  \BibitemOpen
  \bibfield  {author} {\bibinfo {author} {\bibfnamefont {P.~C.}\ \bibnamefont {Rijo}}, \bibinfo {author} {\bibfnamefont {E.~J.}\ \bibnamefont {Vega}}, \bibinfo {author} {\bibfnamefont {F.~J.}\ \bibnamefont {Galindo-Rosales}}, \ and\ \bibinfo {author} {\bibfnamefont {J.~M.}\ \bibnamefont {Montanero}},\ }\bibfield  {title} {\enquote {\bibinfo {title} {On the electrohydrodynamic jet printing of two-dimensional material-based inks for printed electronics},}\ }\href@noop {} {\bibfield  {journal} {\bibinfo  {journal} {Phys. Fluids}\ }\textbf {\bibinfo {volume} {36}},\ \bibinfo {pages} {112014} (\bibinfo {year} {2024})}\BibitemShut {NoStop}%
\bibitem [{\citenamefont {van~der Walt}\ \emph {et~al.}(2012)\citenamefont {van~der Walt}, \citenamefont {Hulsen}, \citenamefont {Bogaerds}, \citenamefont {Meijer},\ and\ \citenamefont {Bulters}}]{WHBMB12}%
  \BibitemOpen
  \bibfield  {author} {\bibinfo {author} {\bibfnamefont {C.}~\bibnamefont {van~der Walt}}, \bibinfo {author} {\bibfnamefont {M.~A.}\ \bibnamefont {Hulsen}}, \bibinfo {author} {\bibfnamefont {A.~C.~B.}\ \bibnamefont {Bogaerds}}, \bibinfo {author} {\bibfnamefont {H.~E.~H.}\ \bibnamefont {Meijer}}, \ and\ \bibinfo {author} {\bibfnamefont {M.~J.~H.}\ \bibnamefont {Bulters}},\ }\bibfield  {title} {\enquote {\bibinfo {title} {Stability of fiber spinning under filament pull-out conditions},}\ }\href@noop {} {\bibfield  {journal} {\bibinfo  {journal} {J. Non-Newtonian Fluid Mech.}\ }\textbf {\bibinfo {volume} {175-176}},\ \bibinfo {pages} {25--37} (\bibinfo {year} {2012})}\BibitemShut {NoStop}%
\bibitem [{\citenamefont {Montanero}\ and\ \citenamefont {Ga{\~n}\'an-Calvo}(2020)}]{MG20}%
  \BibitemOpen
  \bibfield  {author} {\bibinfo {author} {\bibfnamefont {J.~M.}\ \bibnamefont {Montanero}}\ and\ \bibinfo {author} {\bibfnamefont {A.~M.}\ \bibnamefont {Ga{\~n}\'an-Calvo}},\ }\bibfield  {title} {\enquote {\bibinfo {title} {Dripping, jetting and tip streaming},}\ }\href@noop {} {\bibfield  {journal} {\bibinfo  {journal} {Rep. Prog. Phys.}\ }\textbf {\bibinfo {volume} {83}},\ \bibinfo {pages} {097001} (\bibinfo {year} {2020})}\BibitemShut {NoStop}%
\end{thebibliography}

%

\end{document}